Spontaneous organizations of diverse network structures in coupled logistic maps with a delayed connection change

Amika Ohara[1], Masashi Fujii[1], Akinori Awazu[1,2*]

[1]Graduate School of Integrated Sciences for Life Hiroshima University.
[2]Research Center for the Mathematics on Chromatin Live Dynamics, Hiroshima University.
*Corresponding author: Akinori Awazu, awa@hiroshima-u.ac.jp

Abstract
In this study, we performed comprehensive morphological investigations of the spontaneous formations of effective network structures among elements in coupled logistic maps, specifically with a delayed connection change. Our proposed model showed ten states with different structural and dynamic features of the network topologies. Based on the parameter values, various stable networks, such as hierarchal networks with pacemakers or multiple layers, and a loop-shaped network were found. We also found various dynamic networks with temporal changes in the connections, which involved hidden network structures. Furthermore, we found that the shapes of the formed network structures were highly correlated to the dynamic features of the constituent elements. The present results provide diverse insights into the dynamics of neural networks and various other biological and social networks.

I. INTRODUCTION

In biological and social systems, various phenomena are regulated by their self-organized network structures. Neural networks, metabolic networks, food webs, and human communities are typical well-known systems that exhibit self-regulation of network structures through learning, cell differentiation and adaptation, evolution, and communication [1–8]. Such systems have been studied using models that consist of dynamic elements involving changes in the mutual relationships among the elements.

For example, to consider the various properties of neural networks, each element that imitates a nerve cell had to be described as an excitable or chaotic oscillator, and the dynamics of element coupling were assumed to follow a rule inspired by Hebb's law or spike-timing-dependent plasticity (STDP) [9-20]. Recently, it has been suggested that coupled chaotic map systems with Hebbian-like rules may result in systems that could exhibit spontaneous hierarchical network structures with asymmetric couplings, as well as the emergence of pacemakers [21,22].

Such hierarchical structure formations, with various types of pacemaker-like elements, have also been found in various other social systems inspired models [23-25]. However, the universal features and diversity of such self-organized network structures have not been sufficiently investigated.

In this study, we focus on the behaviors of a simple coupled chaotic map system that involves temporal coupling changes inspired by Hebb's rule [26], with a time delay proposed by Ito and Ohira [21]. This model was proposed as one of the simplest models of dynamical network systems that exhibits an obvious hierarchical network structure with a pacemaker element. Furthermore, we found that the simple extension of this model involved the potential to spontaneously form a wider variety of network structures, as mentioned below. In the following arguments, we consider the morphology of self-organized network structures and reveal the relationships between the formed network structures and dynamic features of each element in this model.

## II. MODEL

We consider a simple extension of the Ito-Ohira model [21], which is a coupled logistic map system with a temporal change in the connections (couplings), among elements described by

$$x_{n+1}^i = f((1-c)x_n^i + c \sum_{j=0}^{N-1} w_n^{ij} x_n^i) \quad (1)$$

$$w_{n+1}^{ij} = \frac{[1+\delta \cos \pi (x_n^i - x_{n-1}^j)] w_n^{ij}}{\sum_{j=0}^{N-1}[1+\delta \cos \pi (x_n^i - x_{n-1}^j)] w_n^{ij}} \quad (2)$$

$$f(x) = ax(1-x), \quad (3)$$

where $x_n^i$ $(0 \leq x_n^i \leq 1)$ and $w_n^{ij}$ $(0 \leq w_n^{ij} \leq 1)$ are the state of the $i$-th element and the connection strength from the $j$-th to $i$-th element at time $n$, respectively. $N$ indicates the number of elements $(i,j = 0, ..., N-1)$, and c $(0 \leq c \leq 1)$ indicates the influence strength from the other elements on the dynamics of the $i$-th element. We also consider $\delta$ $(0 \leq \delta \leq 1)$, which is the extension of the model in this study and indicates the connection plasticity; a larger $\delta$ exhibits more sensitive changes in the connection strength. Note that the present dynamical system fully involves the original Ito-Ohira model, and corresponds when $\delta = 1$ [21]. In this model, $w^{ii} = 0$ (no self-connecting) was assumed, and $\sum_j^N w_n^{ij} = 1$ always satisfied for $i$.

For the dynamics of $w_n^{ij}$, the present model employs a simple extension of Hebb's rule [26] where the state of the $j$-th element influences the connection from the $j$-th to $i$-th elements with a one-step time delay. Furthermore, $w_{n+1}^{ij}$ tends to be large when $x_n^i$ and $x^j{}_{n-1}$ are within close proximity to each other. Moreover, the "normalization" of connections strengths over all elements, $\sum_j^N w_{n+1}^{ij} = 1$, can be regarded as a simple representation of the global competition among the connection strength.

In the following simulations, we consider the case $N = 30$ ($i = 0, 1, 2, \ldots, 29$) as in the recent study by Ito and Ohira. For the initial condition of $x_n^i$ and $x_{n-1}^i$, we set $x_n^i = x_{n-1}^i = x_{init}^i$ when $n = 0$, where $x_{init}^i$ is chosen randomly from $x_{init}^i \in (0, 1)$ with uniform probability (we reorder the indices of all elements $i$ to satisfy $x_{init}^0 < x_{init}^1 < \cdots < x_{init}^{28} < x_{init}^{29}$.). We also consider $w_0^{ij} = \frac{1}{N-1}$ ($i \neq j$), assuming that $w_0^{ij}$ is uniformly connected to all the elements, except itself.

We performed simulations of the model for each set of $a$, $c$ and $\delta$ from 20 different initial conditions and focused on the most frequently obtained type of network structures at $n = 10^6 \sim 10^7$ as the typical network structure. Thus, we confirmed that $n = 10^6$ is sufficiently large for the relaxation to the state with its typical network structure for each parameter set. This is supported by the autocorrelation functions of the connection strength among the elements, as discussed below and in Appendix. We classified the typical network structures obtained by various parameter sets according to their connection profiles and the temporal changes in these profiles through the following procedures.

In the present model, the set $w_n^{ij}$ represents the directional network structures of the elements at time $n$. Here, the connection from the $j$-th to $i$-th element is considered strong (weak) when $w_n^{ij}$ is large (small). Now, similar to the recent studies by Ito and Ohira, etc. [9, 10, 21,22], we define the connection from $j$-th to $i$-th elements as existing at time $n$ if $w_n^{ij} > w_a = \frac{1}{N-1}$, where $w_a$ indicates the average of $w_n^{ij}$ over the entire set of $i$ and $j$ ($i \neq j$). Therefore, we classified the connection profiles of typical networks obtained via the proposed model according to this definition.

The dynamic features of connection profiles in typical networks were estimated by the autocorrelation functions of $N(N-1)$ the dimensional vector $W(n) = (\{w_n^{ij}\}_{i \neq j})$ ($n \geq 10^6$) defined as

$$C(\tau) = \frac{<(W(n+\tau)-<W>)(W(n)-<W>)>}{<(W(n)-<W>^2>} \qquad (4)$$

where $<W>$ was assumed as $<W> = \left(\frac{1}{N-1}, \ldots, \frac{1}{N-1}\right)$ ($N = 30$) because the average of $w_n^{ij}$ was estimated to be $\frac{1}{N-1}$ if $w_n^{ij}$ changed ergodically in $n$.

## III. RESULTS

### A. Phase diagram and typical dynamical properties of network structures.

Figure 1 shows the phase diagrams of the typical network structures among the elements for , $a = 3.6, 3.65, 3.7, \ldots, 4.0$, and $c = 0.1, 0.2, \ldots 1.0$ with (a) $\delta = 1$, (b) $\delta = 0.1$, and (c) $\delta = 0.01$. Based on the abovementioned classification, we obtained ten types of qualitatively different states with different structural and dynamic connection network features depending on $a$, $c$ and $\delta$. Each symbol in FIG. 1 indicates the different states, as mentioned in Table 1.

Notably, in the regions of $a$, $c$ and $\delta$, as shown by the black circles in FIG. 1, we obtained $\{x_n^i\}_i$, which showed chaotic and synchronized motions, as seen in FIG. 2 (a). In this case, $w_n^{ij} \to \frac{1}{N-1}$ with $n \to \infty$ indicated that the connections among the elements became uniform, and no specific network structures were formed. In the regions of $a$, $c$ and $\delta$, as shown by the blue squares in FIG. 1, $C(\tau)$ drops to zero for finite $\tau$ as seen in FIG. 2 (b) and Appendix, indicating that all $w_n^{ij}$ changes occurred ergodically in $n$ and no stable network structures appeared.

On the other hand, in the regions of $a$, $c$ and $\delta$, as shown by the orange triangles, yellow circles, yellow standing triangles, red diamonds, or red stars in FIG. 1, $C(\tau) \sim 1$ for all $\tau$ (Appendix), indicated that the network structures were stable and unchanged once they were formed. We call these stable networks. We also found that, in other cases, $C(\tau)$ decreased but for a relaxed finite value for $\tau \to \infty$ (Appendix). This indicated that the networks of these observed states are not time stable, but hidden basic network structures do exist and $w_n^{ij}$ changes temporally around such basic structures. We call these dynamic networks.

Based on these facts, we next focused on the structural and dynamical features of the typical stable and dynamic network structures, where $C(\tau) > 0$ for $\tau \to \infty$, as described by $\{w_n^{ij}\}_{i \neq j}$, and the dynamic properties of elements, as described by $\{x_n^i\}_i$.

### B. Morphology of typical structures of stable networks

We considered the morphology of typical states with stable network structures that $C(\tau) \sim 1$ are maintained for $\tau \to \infty$. To characterize the structural properties of observed networks, we defined two values, the indegree and outdegree of the $i$-th element at time $n$, $ID_n^i$ and $OD_n^i$, defined by the number of connections to and from the $i$-th element, as calculated by $ID_n^i = \sum_j^N \theta(w_n^{ij} - \frac{1}{N-1})$ and $OD_n^i = \sum_j^N \theta(w_n^{ji} - \frac{1}{N-1})$, respectively. Here, $\theta(x)$ is the Heviside function that yields $\theta(x) = 1$ for $x > 0$ and $\theta(x) = 0$ for $x \leq 0$. Notably, this is because the typical networks obtained below are stable; $ID_n^i$ and $OD_n^i$ converge to the respective constant values $\to ID^i$ and $\to OD^i$ for sufficiently large $n$.

In the region marked by orange triangles in FIG. 1 ($a = 3.65, 3.7, c = 0.2, \delta = 1.0, 0.1$ and $a = 3.8, c = 0.2, \delta = 0.01$), we found the formations of network structures named "Pacemaker network" at sufficiently large $n$ values, as shown in FIG. 3. In these states, one or a few elements exhibit, $OD^i = N - 1$ but most elements exhibit $OD^i = 0$. On the other hand, the $ID^i$ of all the elements were equal to the number of elements with $OD^i = N - 1$. We called these few elements "pacemaker(s)" because $x_n^i$ they influence the $x_{n+1}^i$ of the other elements effectively, but those of most other elements do not.

These states, moreover, $\{x_n^i\}_i$ show the chaotic motions and $x_n^i$ of some of the elements that are synchronized with pacemakers, but the $x_n^i$ of the elements that are desynchronized with pacemakers also synchronized with each other, as shown in FIG. 3 (c-d).

The typical number of pacemakers depends on the parameter set: i) only one pacemaker appears when $a = 3.7, c = 0.2, \delta = 1.0$ or $0.1$ ii) two pacemakers appear when $a = 3.65, c = 0.2, \delta = 1.0$ iii) two to five pacemakers appear depending on the initial conditions when $a = 3.65, c = 0.2, \delta = 0.1$, $a = 3.65, c = 0.2, \delta = 0.01$, and $a = 3.7, c = 0.2, \delta = 0.01$ as shown in FIG. 4. Note that in case i), $\delta = 1$ is the same as a result reported by Ito and Ohira [21].

In the region marked by yellow circles in FIG. 1 ($a = 3.6 \sim 3.8, c = 0.1, \delta = 1.0, 0.1, 0.01$ and $a = 3.85, c = 0.1, \delta = 0.1, 0.01$), we found the formations of the network structures named "paired layers network" were sufficiently large, and $n$, in which the elements are divided into paired layers, and each element is connected to and from other elements that belong to another layer, is shown in FIG. 5 (a-c). Here, $\{x_n^i\}_i$ shows the periodic motions, which are divided into two clusters, where $x_n^i$ represents the elements in the same synchronized layer. This is shown in FIG. 5 (c-d). In these states, the indegree and outdegree of each element differ according to $ID^i < OD^i$ for some elements but $ID^i \gg OD^i$ is often seen for others, as shown in FIG. 5 (e-f).

In the region highlighted by red diamonds in FIG. 1 ($a = 3.9, c = 0.1, \delta = 0.1, 0.01$), we found the formations of network structures called the "loop network" at sufficiently larger $n$ values, in which the elements were divided into four layers, as shown in FIG. 6 (a–c). Here, the $\{x_n^i\}_i$ of the elements showed periodic motions, and the $x_n^i$ of the elements in the same layer were synchronized, as shown in FIG. 6 (c-d). In these network structures, the elements in the same layer were commonly connected from all the elements belonging to one of the three other layers, and were commonly connected to all elements belonging to another of two other layers, and then four layers to form a loop.

In the region marked by yellow standing triangles in FIG. 1 ($a = 3.85, 3.9, c = 0.1, \delta = 1.0$), we found network structure formations named "multi-layers network" at a sufficiently large $n$, in which the elements were divided into more than three layers, and each element was connected to and from other elements that belong to other layers, as shown in FIG. 7 (a-c). Here, $\{x_n^i\}_i$ the periodic motions and $x_n^i$ of the elements in the same layer were synchronized, as shown in FIG. 7 (c-d). In these cases, the observed networks contained various types of layers such as those consisting of elements with $ID^i < OD^i$, those consisting of elements with $ID^i > OD^i$, and elements with few or zero $OD^i$, where four layers form a similar network structure to the loop network. Moreover, such networks also contain multiple hierarchies with the elements in the upper four layers forming a loop and lower layers exhibiting asymmetric influence flows among the elements.

In the region highlighted by red stars in FIG. 1 ($a = 3.95, 4.0, c = 0.1, 0.2, \delta = 1.0, 0.1, 0.01$), we found that distinct small hierarchical networks with upstream and downstream elements had formed at sufficiently large $n$ values, as shown in FIG. 8 (a-c). Here, the $\{x_n^i\}_i$ of the elements showed chaotic motions, and the $x_n^i$ of the elements were not synchronized even if they belonged to the same networks, as shown in Fig. 8 (d). In these networks, each distinct network resembled a small "pacemaker network," but the connections of each network were sparser than the abovementioned pacemaker networks.

C. Morphology of typical structures of dynamic networks with hidden structures

Next, we considered the morphology of the typical states where $C(\tau)$ decreases in $\tau$ but $C(\tau) > 0$ were maintained for $\tau \to \infty$. In these cases, as mentioned below, we found various temporal changes in the shapes of the connected network. Still, these changes seemed to be restricted to various parameter-dependent hidden structures. To evaluate the structural and dynamic features of the typical networks for each parameter set, we focused on the temporal changes in indegrees and outdegrees, $ID_n^i$ and $OD_n^i$, and long-term indegrees and outdegrees, $LID^i$ and $LOD^i$. The latter was estimated by $LID^i = \sum_j^N \theta(\langle w_n^{ij} \rangle - \frac{1}{N-1})$ and, $LOD^i = \sum_j^N \theta(\langle w_n^{ji} \rangle - \frac{1}{N-1})$ where $\langle w_n^{ij} \rangle$ indicates the average $w_n^{ij}$ over n = $8 \times 10^6 \sim 10^7$.

In the region marked by white squares in FIG. 1, ($a = 3.75, 3.8, c = 0.2, \delta = 1, 0.1, 0.01$, $a = 3.7, 3.75, 3.8, 3.85, c = 0.3, \delta = 1, 0.1, a = 3.9, c = 0.4, \delta = 1$ and $a = 4.0, c = 0.5, \delta = 1$), we found that the states exhibited a dynamic network with hidden structures called the "hidden paired layers network" at sufficiently large $n$ values, as shown in FIG. 9. Here, a group of elements involving greater $OD_n^i$ compared to the others appears for the upstream elements in hierarchical networks or pacemaker networks, as mentioned previously, as shown in FIG. 9 (a–c). However, such a network is not stable but transits temporally. In this state, the $\{x_n^i\}_i$ of the elements exhibited chaotic motions, as shown in FIG. 9 (d). After a certain amount of time, the elements that played upstream element roles with large $OD_n^i$ replaced the role with other elements, as shown in FIG. 9 (a–c). Here, the lifetime order of the tentative networks was expected to be $\tau$ to $C(\tau)$ as shown by the relaxed constant in FIG. 9 (f).

As a result of such temporal connection changes among elements, in the cases of $= 3.75, 3.8, c = 0.2, \delta = 1, 0.1, 0.01$, $a = 3.7, 3.75, 3.8, 3.85, c = 0.3, \delta = 1, 0.1$ and $a = 3.9, c = 0.4, \delta = 1$, the long averaged time of the network structures in such cases involved two groups of elements with higher and lower $LOD^i$, as shown in FIG. 9 (g-h).

Thus, this seems similar to the "paired layers network" introduced previously. On the other hand, in the cases of $a = 4.0, c = 0.5, \delta = 1$, the long averaged time of the network structures in such case exhibited similar to the "pacemaker network" also introduced previously, as shown in FIG. 10.

In the region marked by the yellow laying triangle in FIG. 1 ($a = 3.9, c = 0.2, \delta = 0.01$), we found that the states exhibited a dynamic network with hidden structures called the "hidden modular network" at sufficiently larger $n$ values, as shown in FIG. 11. Here, most elements were divided into modules consisting of four elements. In each $n$, the four element forming module sets were strongly connected with each other, and some modules were connected to the elements that did not belong to any module. In contrast, others existed individually, as shown in FIG. 11 (a-c). Here, the $x_n^i$ of the elements exhibited desynchronized chaotic motions even for the same modules, as shown in Fig. 11 (d). However, the construction of each module was unchanged. Notably, elements that did not belong to any modules were connected from all elements on average, as shown in FIG. 11 (g-h), while such connections change temporally. Moreover, as shown in FIG. 12, the order of the standard deviation of each connection was comparable to its average value, and the connections among elements also changed temporally even in the same modules. Still, each set of four elements in each module was unchanged.

In the region marked by the white square in FIG. 1 ($a = 3.95, c = 0.2, \delta = 1.0, 0.1, 0.01$), we found that the states exhibited a dynamic network with hidden structures called the "hidden randomly connected network" at sufficiently large $n$ values, as shown in FIG. 13. Here, the $\{x_n^i\}_i$ of the elements showed chaotic motions, and various complex networks where elements seem to be connected randomly appeared tentatively. The average order of the lifetime of each tentative network structure was estimated by $\tau$, to which $C(\tau)$ relaxes the constant in FIG. 13 (f). However, some $\langle w_n^{ij} \rangle$ remained larger than $1/(N-1)$. Each element involved five to ten of $LID^i$ and $LOD^i$ with, $LID^i \sim LOD^i$ as shown in FIG. 13 (g-j), by which the system exhibited hidden network structures with (not sparse but also not dense) random connections among the elements without any pacemaker-like specific elements or significant directionalities.

D. Relationships between formed network structures and the dynamics of the elements
Then, we focused on the relationship between the formed network structures and dynamics of $\{x_n^i\}_i$.

To estimate such relations, we calculated the split exponent (tangential Lyapunov exponent) of the $i$-th element [27] as

$$\lambda^i = \ln(1-c) + \frac{1}{T}\lim_{T\to\infty}\sum_{n=1}^{T} \ln|f'(x_n^i)| \qquad (5)$$

$$f(x) = ax(1-x). \qquad (6)$$

We focused on the maximum split exponent $\lambda^{max} = \max\{\lambda^i \mid i = 0, 1, ... N-1\}$ in the following arguments: $\_^{max}$ were measured for $a = 3.6, 3.61, 3.62, \_\_, 4$ and $c = 0, 0.01, 0.02, \_\_, 0.6$ to compare them to the phase diagram of the dynamics of $\{x_n^i\}_i$, as reported in a recent study [21]. Notably, we evaluated the $\lambda^{max}$ using $\{x_n^i\}_i$ for n = 5×10⁶ ~ 10⁷ from five different initial conditions, but the results were almost the same, independent of the initial conditions.

In the case of $\delta = 1.0$ the phase diagram of the $a$ and $c$ dependent dynamics of $\{x_n^i\}_i$ were reported, where their dynamics were classified as the coherent phase, ordered phase, partially ordered phase, and desynchronized phase [21]. As shown in Fig. 14 (a), the boundary curves among these phases in the phase diagram were closely related to the $\lambda^{max}$ of the landscape, as follows. The boundary between the coherent phase and the partially ordered phase was obtained as a curve located at $\lambda^{max} = -0.1 \sim -0.05$ for $c$ larger than ~ 0.2. Furthermore, the boundary between the partially ordered phase and the ordered phase was obtained as a curve, which grew along some contours, and was from $\lambda^{max} \sim 0$ $c$ to between ~ 0.15 and ~ 0.2. Finally, the boundary between the ordered phase and the desynchronized phase was obtained as a curve, which grew along some contours from $\lambda^{max} \sim 0$ at $c < 0.15$.

FIG. 14 (b) shows the superposition of $\lambda^{max}$ and the phase diagram of the network structure in Fig. 1, for the case of $\delta = 1.0$. By comparing Fig. 13 (a) and (b), we found the following correlations between the formed typical network structures and the dynamics of the elements: most regions that exhibited uniform $w_n^{ij}$ (black circles) matched the coherent phase. There were no typical network structures (blue squares) when the dynamics of the elements were partially ordered with larger $a$ than ~3.75, while the pacemaker networks (orange triangles) appeared for smaller $a$. The paired layer networks (yellow circles) and multi-layer networks (yellow standing triangles) were found in cases where the elements showed ordered motions. In the cases where the elements exhibited desynchronized motions, states with distinct hierarchical networks (red stars) appeared in the region near the boundary between the coherent phase and partially ordered phase, and dynamic networks with hidden paired layers network

(white squares) were formed. Furthermore, in the region near the boundary between the desynchronized phase and the ordered motions, dynamic networks with hidden randomly connected networks (purple squares) were formed.

Similar correlations were formed between typical network structures, and the dynamics of elements are also found in the cases of $\delta = 0.1$ and $0.01$. Figure 15 (a-b) shows $\lambda^{max}$ for $\delta = 0.1$ and $0.01$ where three curves are drawn according to the same criteria as those in FIG. 14 (a), and FIG. 15 (c-d) show the superposition of $\lambda^{max}$ and the phase diagram of the network structure in FIG. 1 in these cases. We found almost the same relationships between the formed typical network structures and the dynamic properties of the elements, as shown in FIG. 14 (b), except for the following cases. In the region near the boundary between the two regions corresponding to the desynchronized phase and the ordered phase in Fig. 15 (a) and (b), the loop network structure (red diamonds) was observed differently from the case in $\delta = 1$. Furthermore, in the region near the boundary between the two regions corresponding to the partially ordered motions and the ordered phase in Fig. 15 (a), a dynamic network with a hidden modular network (yellow laying triangle) appeared in the case of $\delta = 0.01$.

## IV. DISCUSSION

In this study, we considered the behaviors of a simple coupled map system involving temporal changes, which had connections among elements with a time delay proposed by Ito and Ohira [21]. We found that the present system has the potential to exhibit a variety of network structures of connections among elements, either as stable structures or hidden structures.

A recently proposed model inspired by human communication employed the rule of connection among elements to emerge as the leader elements (almost the same meaning as the pacemakers in this study), and it exhibited various types of hierarchical network structures [23]. We also found similar hierarchical networks in the present model, although our discussed model did not contain any specific rules to emerge pacemakers explicitly. Since there are no explicit rules for the emergence of pacemakers, the present model exhibited other types of networks, such as loop networks and dynamics networks with hidden modular networks.

The present model employs the extension of Hebb's rule, where the connection from an element to the other at the time tends to be strengthened if the state of the latter at a particular time and that of the former at one time step before are similar. We could regard this rule containing the time delay as one of the simple descriptions of the effects of spike-timing-dependent-plasticity that is observed in neural network systems [28]. Still, we need more detailed qualitative and quantitative considerations. Recently, various dynamical system models with a change in the connections to the simple Hebb's like rules were studied, and various hierarchical networks were observed. Still, the formed network structures were dependent on the properties of the models [24, 25, 29]. On the other hand, our discussed model in the present arguments showed not only various hierarchical networks but also other types of networks that were not model-dependent but also parameter-dependent. We expect that the time delay in the rule of connection changes plays a key role in creating such various network structure formations.

A more detailed parameter-dependent study is still needed that observes the behaviors of the present model in the future since this model is expected to involve further potential, specifically to emerge a richer variety of network structures. We also need to conduct additional studies on the dynamics and functions of neural networks and other biological networks based on the obtained results and modifications of the present model.

## ACKNOWLEDGMENTS

We thank J. Ito for fruitful information provision and H. Nishimori, M. Shiraishi, and H. Takagi for fruitful discussions. This work was supported by JSPS KAKENHI Grant Number 19K20382 (M. F.).

TABLE 1. Summary of classification of typical network structures and dynamics of elements.

| Symbol in FIG. 1 | Network structure | Stability of network | Dynamics of $x_n^i$ | Details in |
| --- | --- | --- | --- | --- |
| Blue □ | None | Unstable | Chaotic | None |
| Black ○ | Uniform | Stable | Chaotic | None |
| Orange ▽ | Pacemaker(s) | Stable | Chaotic | FIG. 3, 4 |
| Yellow ○ | Pared layers | Stable | Periodic | FIG. 5 |
| Red ◇ | Loop | Stable | Periodic | FIG. 6 |
| Yellow △ | Multi-layers | Stable | Periodic | FIG. 7 |
| Red ☆ | Distinct Hierarchical | Stable | Chaotic | FIG. 8 |
| White □ | Hidden Pared layers | Dynamic | Chaotic | FIG. 9, 10 |
| Yellow ▷ | Hidden Modular | Dynamic | Chaotic | FIG. 11, 12 |
| Purple □ | Hidden random | Dynamic | Chaotic | FIG. 13 |

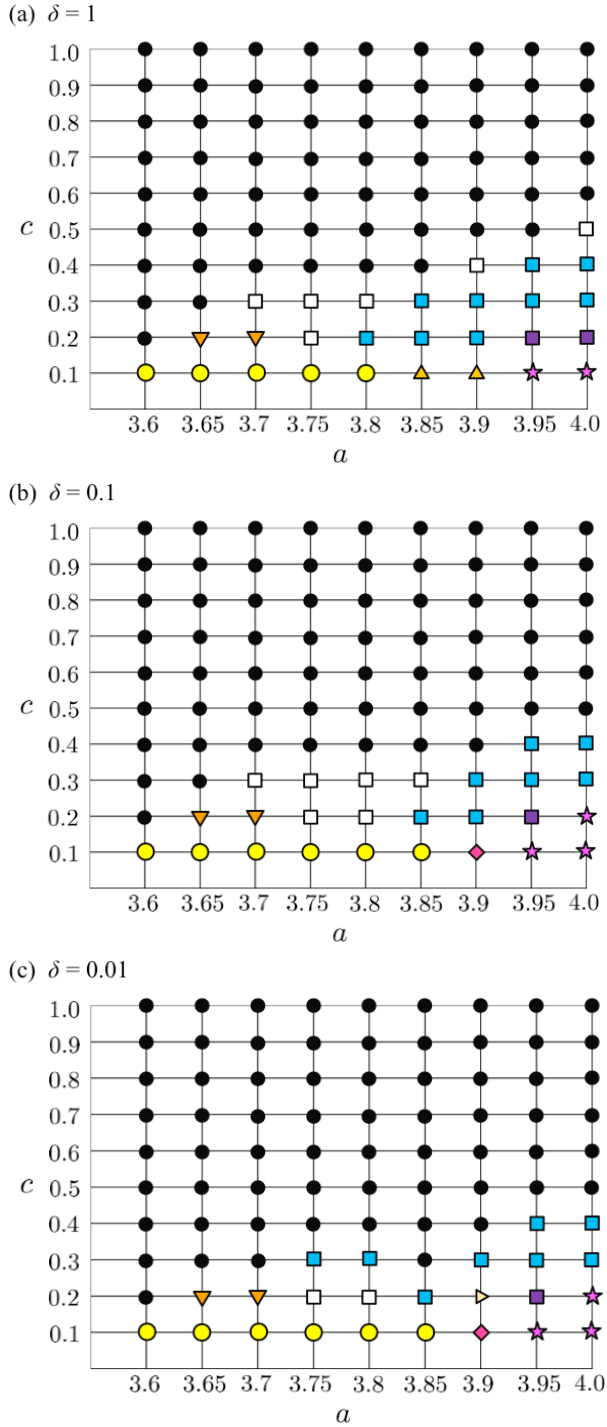

FIG. 1. Phase diagram of the network structures for (a) $\delta = 1$, (b) $\delta = 0.1$, and (c) $\delta = 0.01$ for $a = 3.6 \sim 4.0$, and $c = 0.1 \sim 1.0$. Each symbol represents the mean typical network structures observed in the case of the given parameter set. Details of each symbol are shown in TABLE 1.

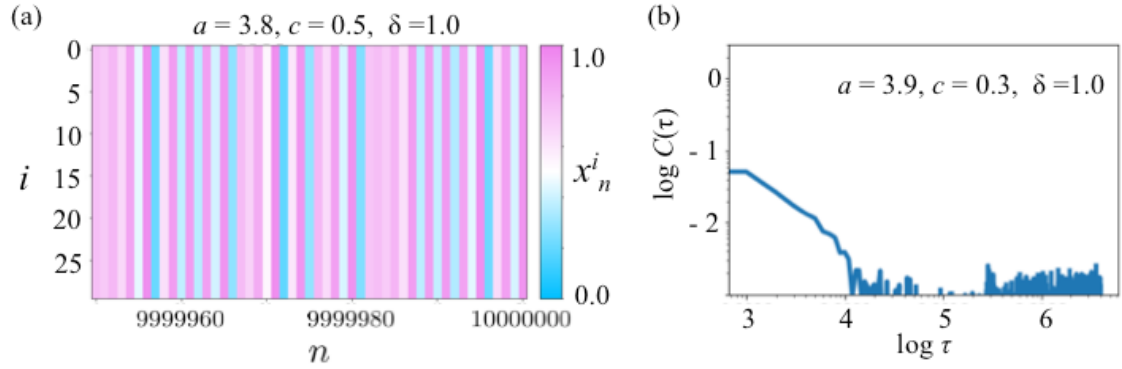

FIG. 2. (a) Typical temporal evolutions of $x_n^i$ of each element in the present model with $a = 3.8, c = 0.5, \delta = 1.0$ for $n = 9999950 \sim 10000000$. (b) Typical autocorrelation function of $\{w_n^{ij}\}_{i \neq j}$ of each element in the present model with $a = 3.9, c = 0.3, \delta = 1.0$.

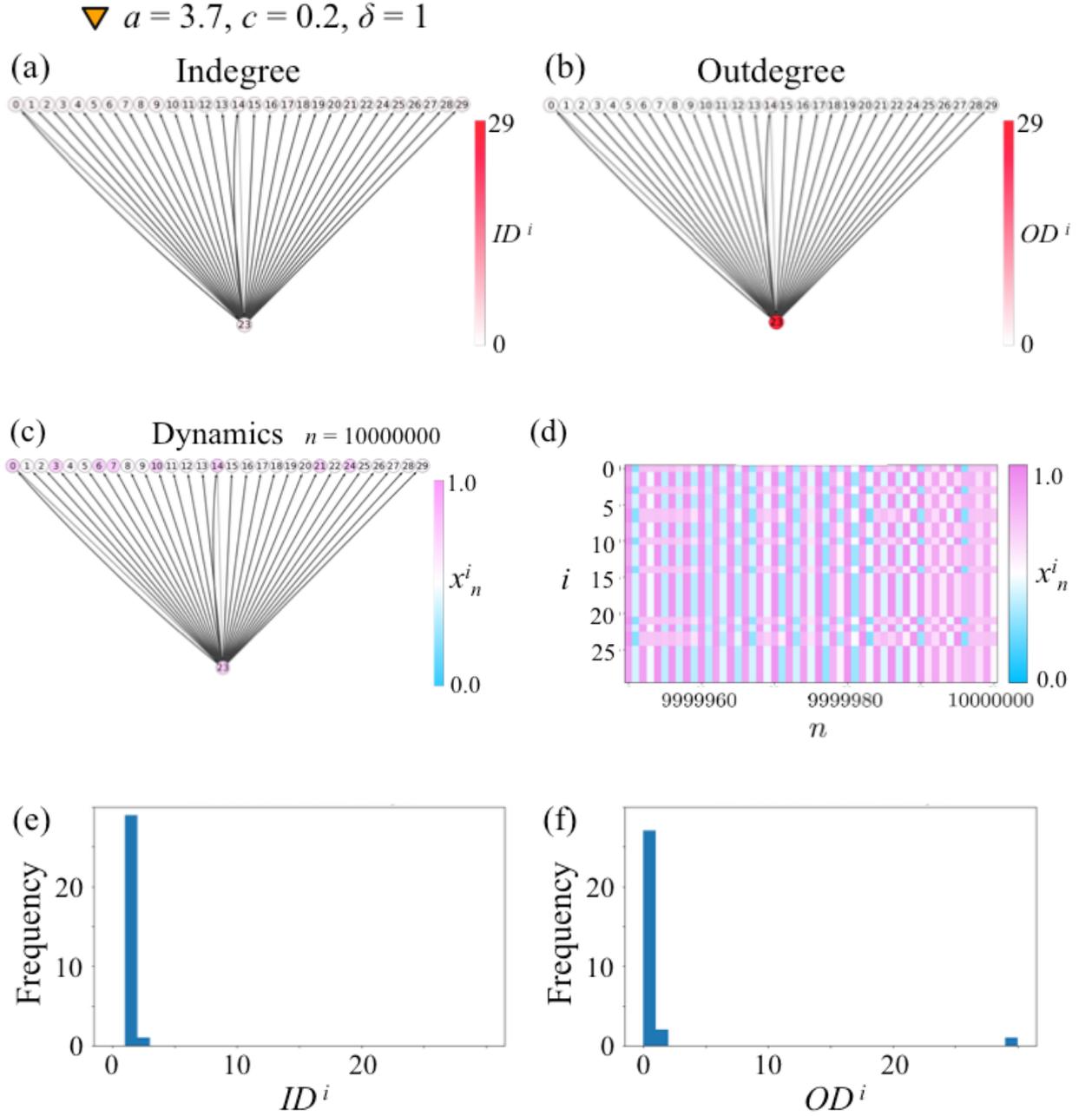

FIG. 3. (a-c) Example of typical network structure, (a) $ID^i$, (b) $OD^i$, and (c) $x_n^i$ at $n = 10^7$ of each element in the present model with $a = 3.7, c = 0.2, \delta = 1.0$. The network structure named "Pacemaker network" was observed in this state. Here, the circle with index $i$ indicates the symbol of $i$-th element, and the color of each indicates the values of (a) $ID^i$, (b) $OD^i$, and (c) $x_n^i$. When $w_n^{ij} > \frac{1}{N-1}$ at $n = 10^7$, the $i$-th element is connected from $j$-th element by an arrow. Notably, $w_n^{ij} > \frac{1}{N-1}$ for $n \geq 10^6$ since the presented network is stable. (d) Typical temporal evolutions of $x_n^i$ for $n = 9999950 \sim 10000000$, and histograms of (e) $ID^i$ and (f) $OD^i$ in the presented state.

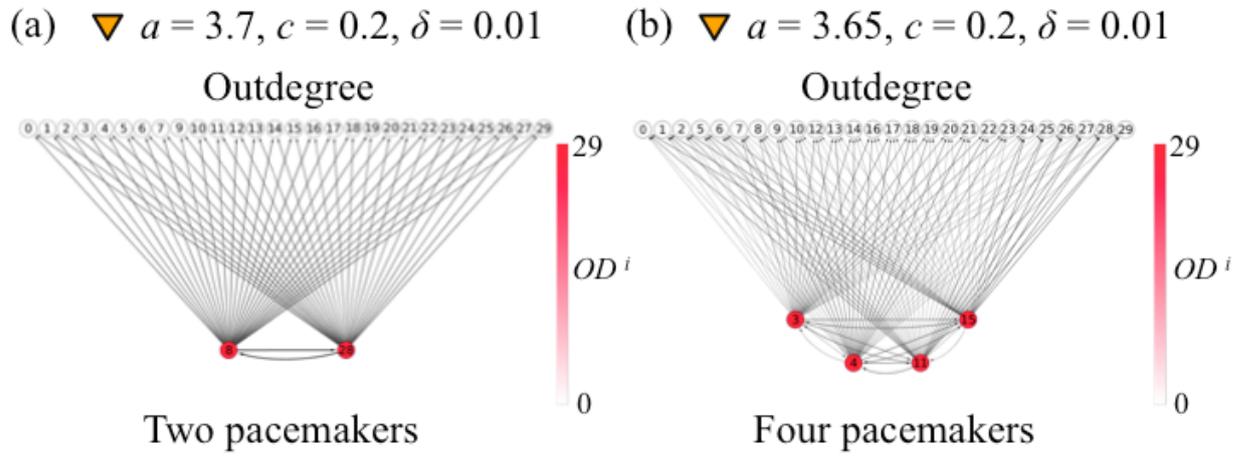

FIG. 4. Example of typical network structure and $OD^i$ of Pacemaker networks with (a) two pacemakers for $a = 3.7, c = 0.2, \delta = 0.01$ and (b) four pacemakers for $a = 3.65, c = 0.2, \delta = 0.01$. The meanings of the circles and arrows are the same as those in FIG. 2.

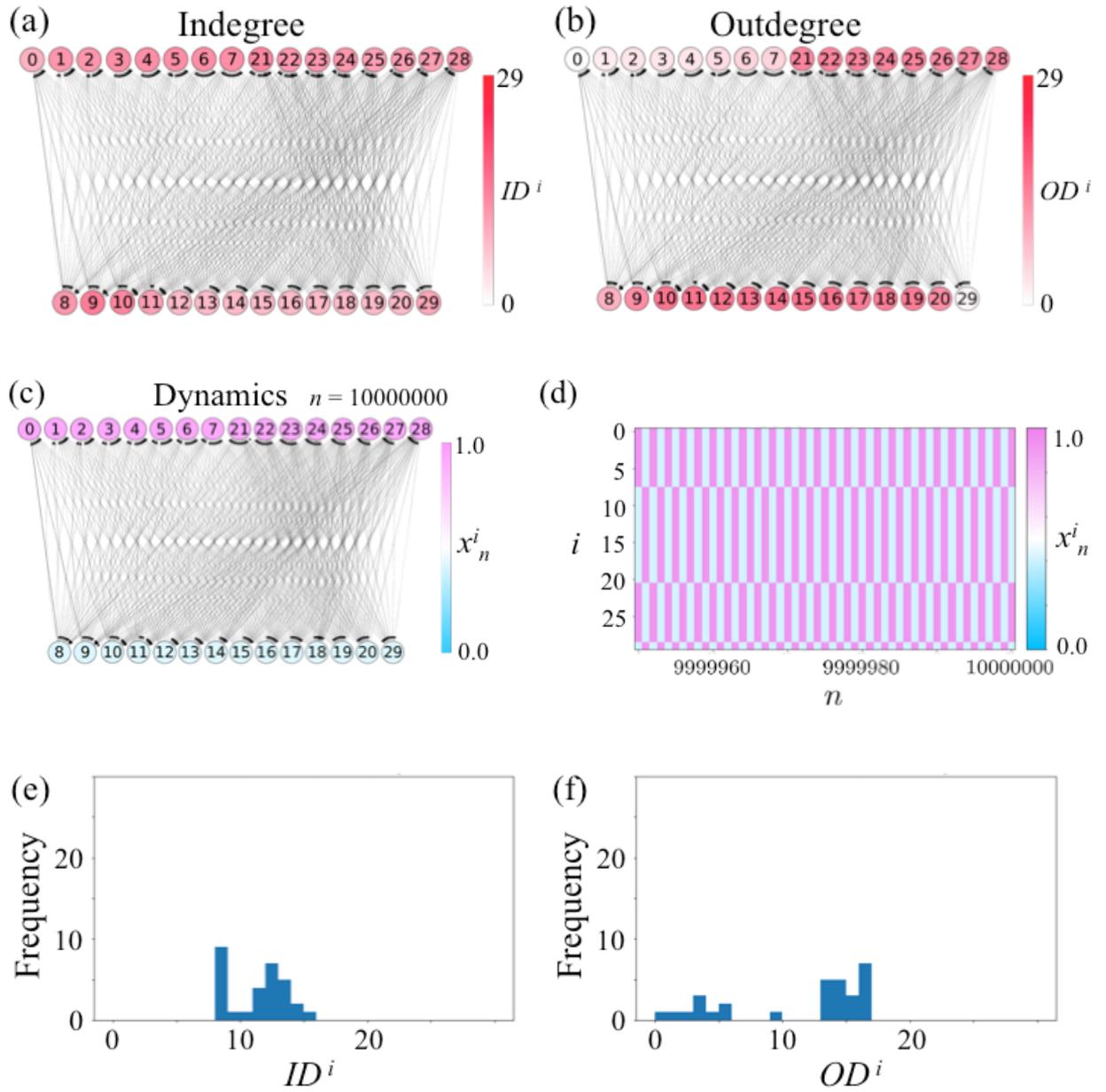

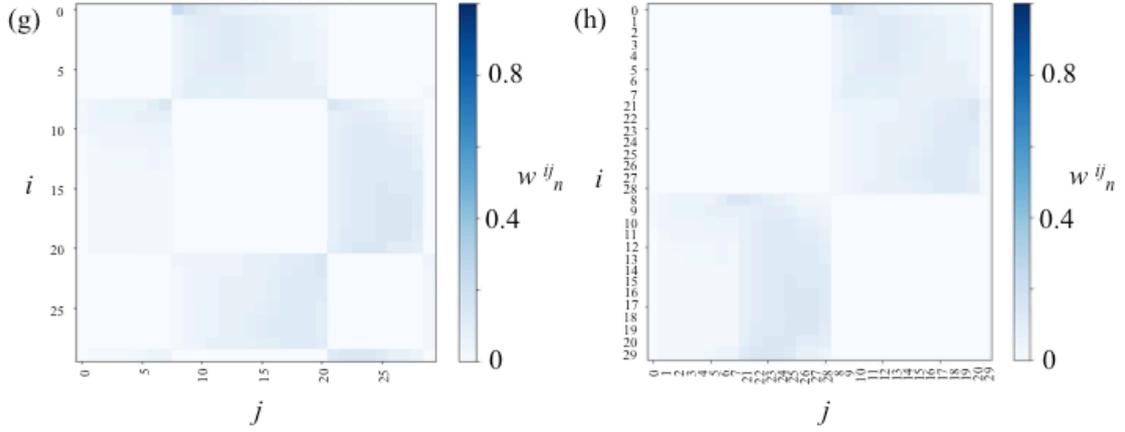

FIG. 5. (a-c) Example of a typical network structure, (a) $ID^i$, (b) $OD^i$, and (c) $x_n^i$ at $n = 10^7$ of each element in the present model with $a = 3.7, c = 0.1, \delta = 1.0$. The network structure called "pared layers network" was observed in this state. The meanings of the circles and arrows are the same as those in FIG. 2. (d) Typical temporal evolutions of $x_n^i$ for $n = 9999950 \sim 10000000$. (e-h) histograms of (e) $ID^i$ and (f) $OD^i$, and (g) $w_n^{ij}$, in the order from small to large $i, j$, and (h) $w_n^{ij}$ for the reordered $i, j$ in the presented state.

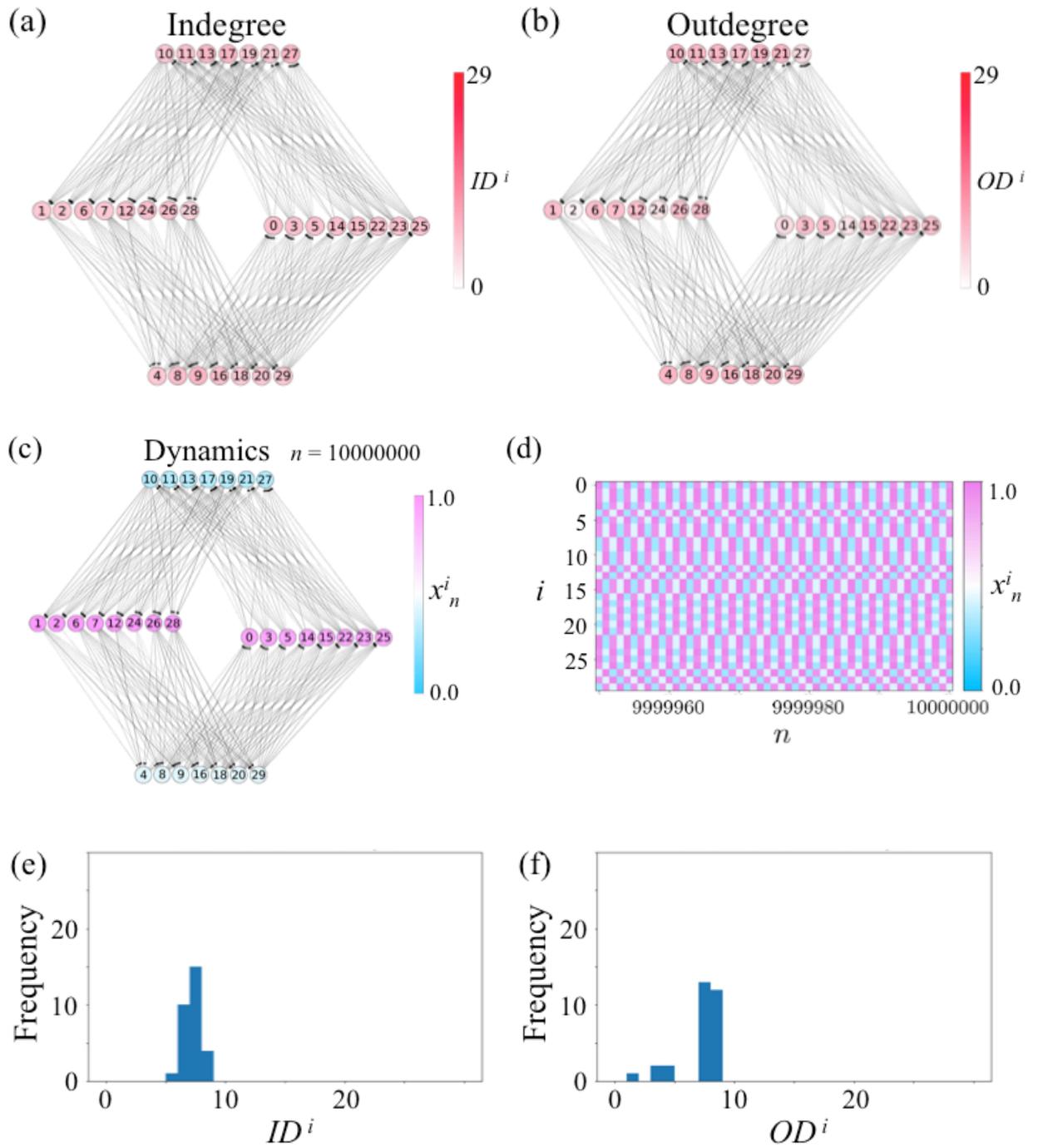

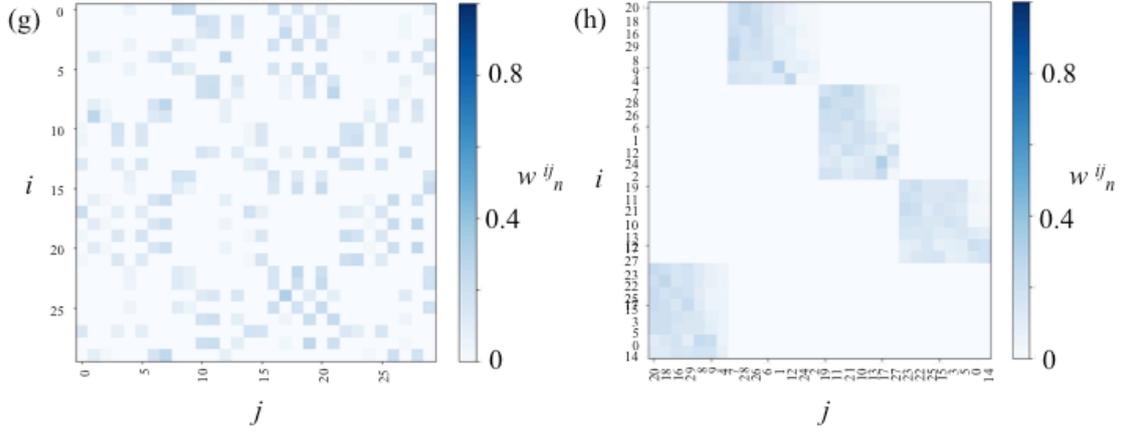

FIG. 6. (a-c) Example of a typical network structure, (a) $ID^i$, (b) $OD^i$, and (c) $x_n^i$ at $n = 10^7$ of each element in the present model with a = 3.9, c = 0.1, δ = 0.1. The network structure called the "loop network" was observed in this state. The meanings of the circles and arrows are the same as those in FIG. 2. (d) Typical temporal evolutions of $x_n^i$ for $n = 9999950 \sim 10000000$. (e-h) histograms of (e) $ID^i$ and (f) $OD^i$, and (g) $w_n^{ij}$, in the order from small to large $i, j$, and (h) $w_n^{ij}$ for the reordered $i, j$ in the presented state.

△ $a = 3.9, c = 0.1, \delta = 1$

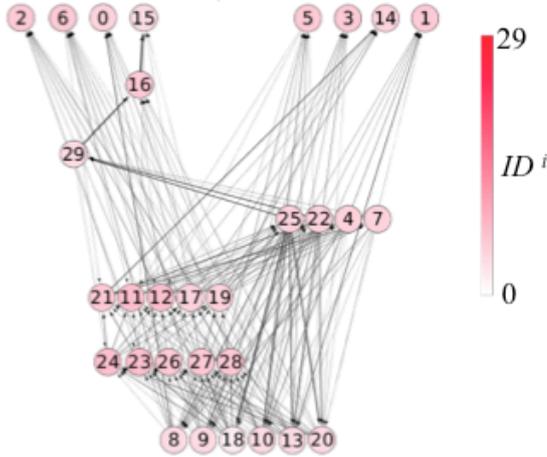
(a) Indegree

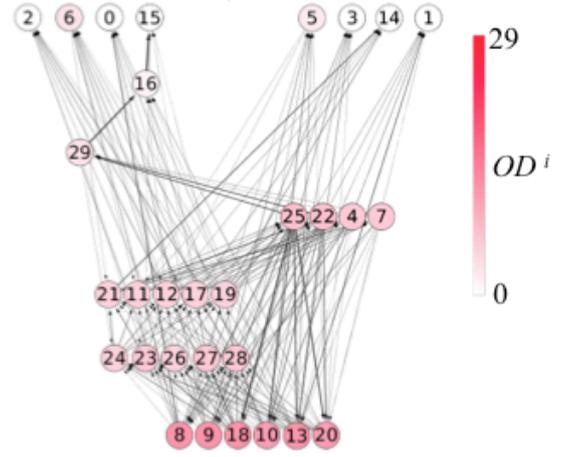
(b) Outdegree

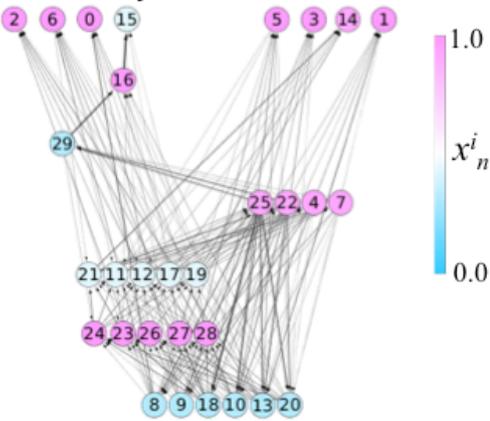
(c) Dynamics $n = 10000000$

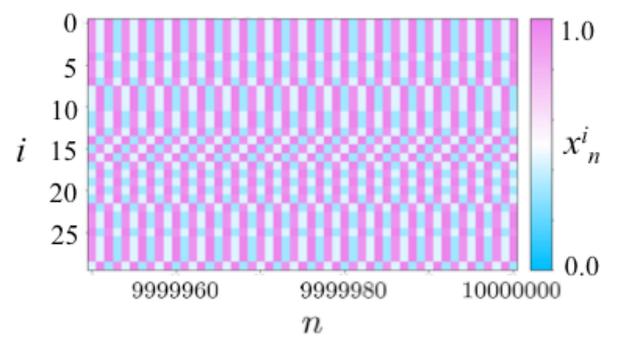
(d)

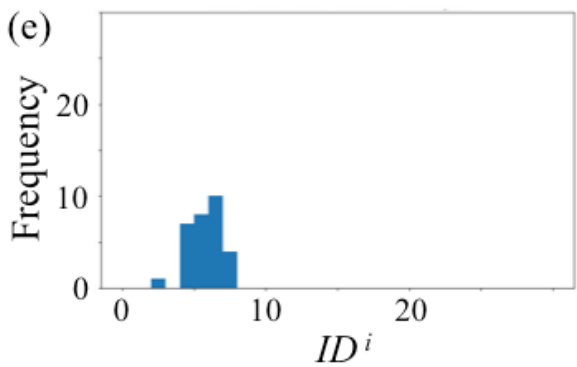
(e)

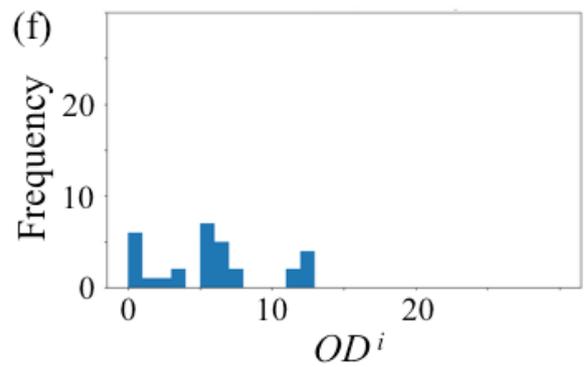
(f)

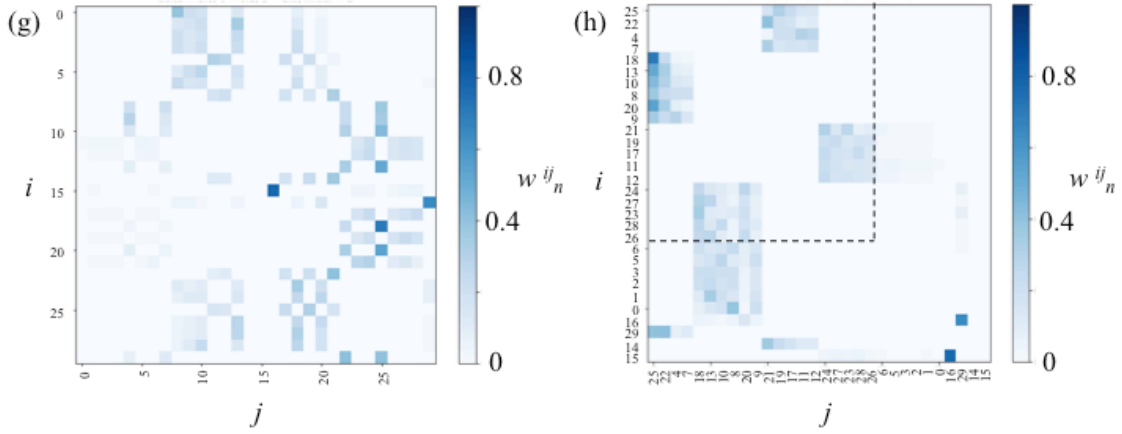

FIG. 7. (a-c) Example of typical network structure, (a) $ID^i$, (b) $OD^i$, and (c) $x_n^i$ at $n = 10^7$ of each element in the present model with $a = 3.9, c = 0.1, \delta = 1.0$. The network structure called "multi-layers network" was observed in this state. The meanings of circles and arrows are the same as those in FIG. 2. (d) Typical temporal evolutions of $x_n^i$ for $n = 9999950 \sim 10000000$. (e-h) histograms of (e) $ID^i$ and (f) $OD^i$, and (g) $w_n^{ij}$, in the order from small to large $i, j$, and (h) $w_n^{ij}$ for the reordered $i, j$ in the presented state. Left upper region in (h) means that a loop structure exists in the upper stream of this network structure.

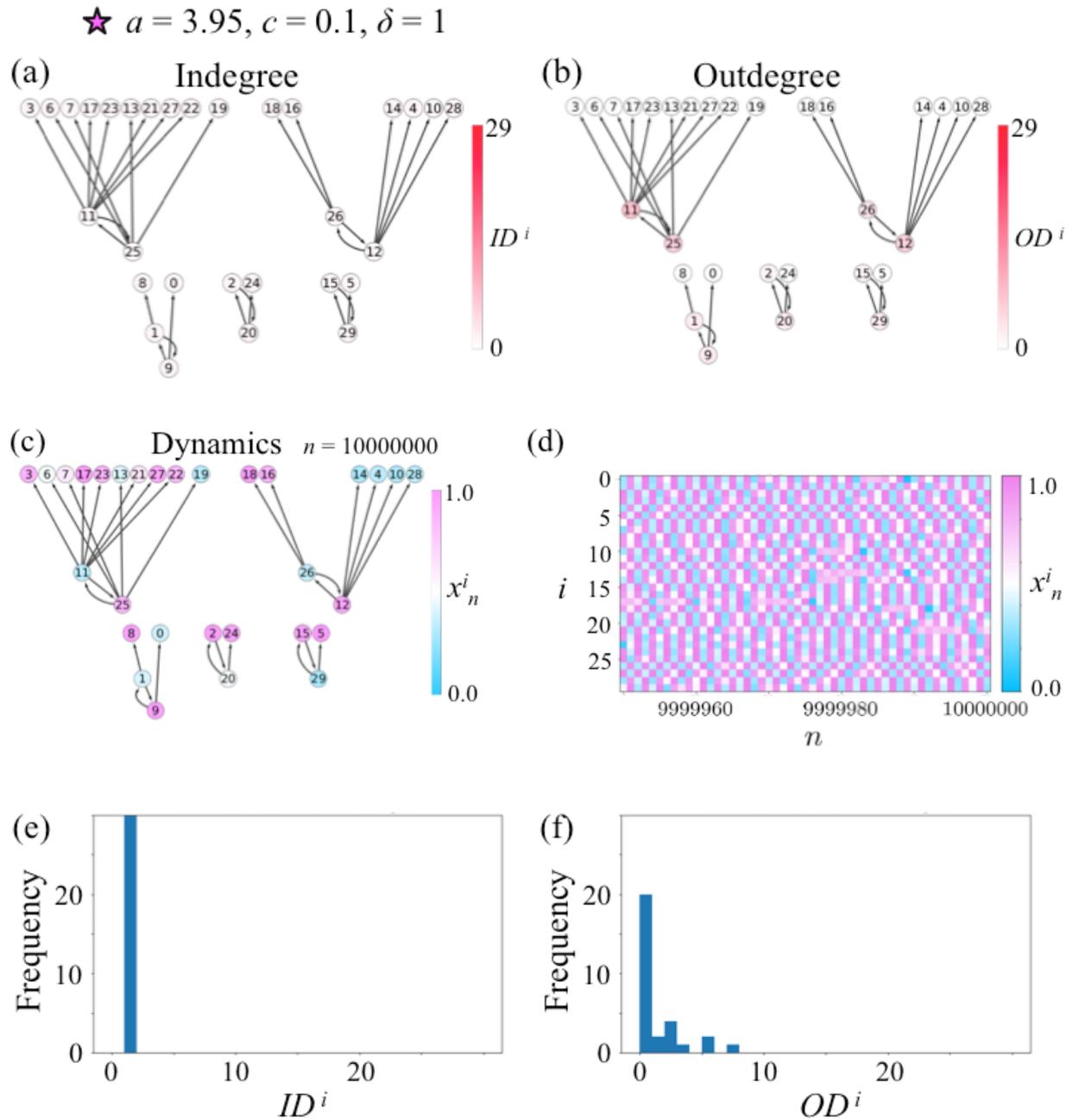

FIG. 8. (a-c) Example of typical network structure, (a) $ID^i$, (b) $OD^i$, and (c) $x_n^i$ at $n = 10^7$ of each element in the present model with a = 3.95, c = 0.1, δ = 1.0. In this case, some distinct hierarchical networks with upstream and downstream elements were observed. The meanings of circles and arrows are the same as those in FIG. 2. (d) Typical temporal evolutions of $x_n^i$ for $n = 9999950 \sim 10000000$ and histograms of (e) $ID^i$ and (f) $OD^i$ are in the presented state.

$a = 3.75, c = 0.3, \delta = 1$

(a) Indegree

(b) Outdegree

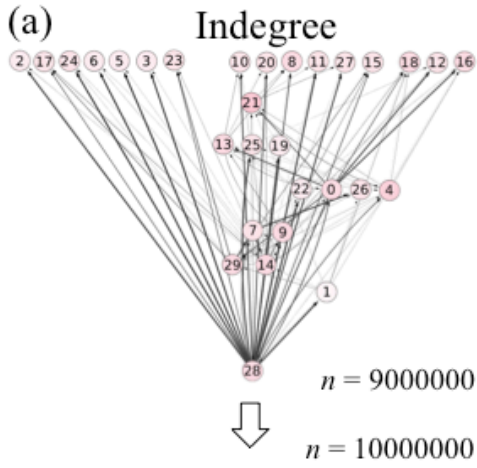
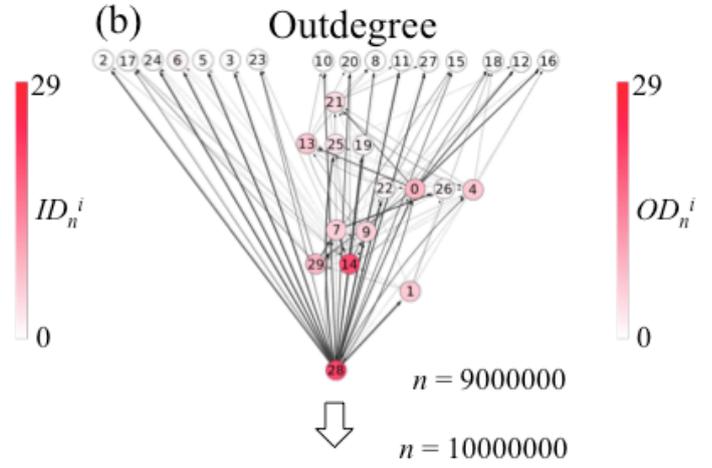

$n = 9000000$

$n = 10000000$

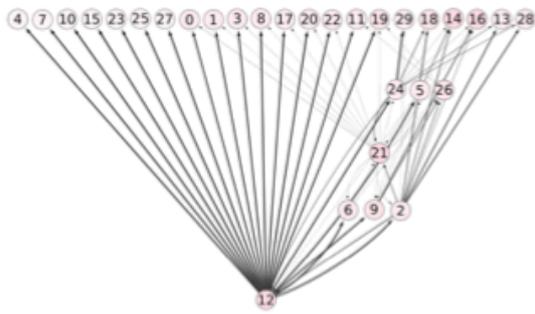
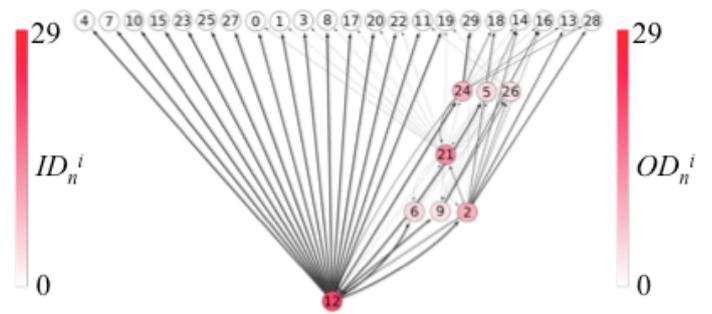

(c) Dynamics

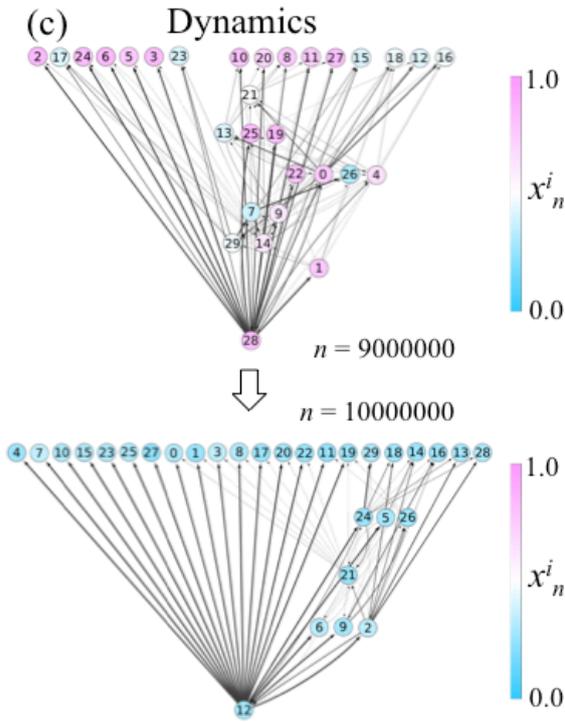

$n = 9000000$

$n = 10000000$

(d)

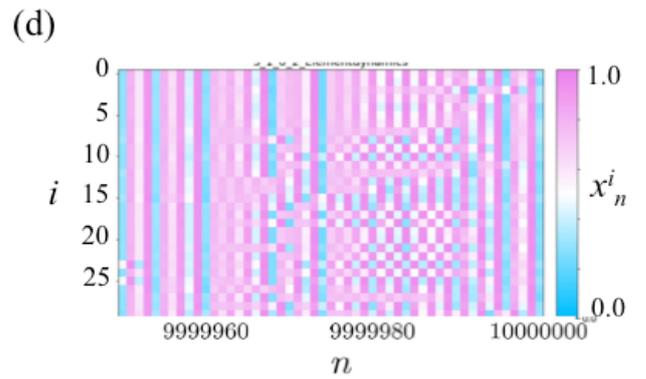

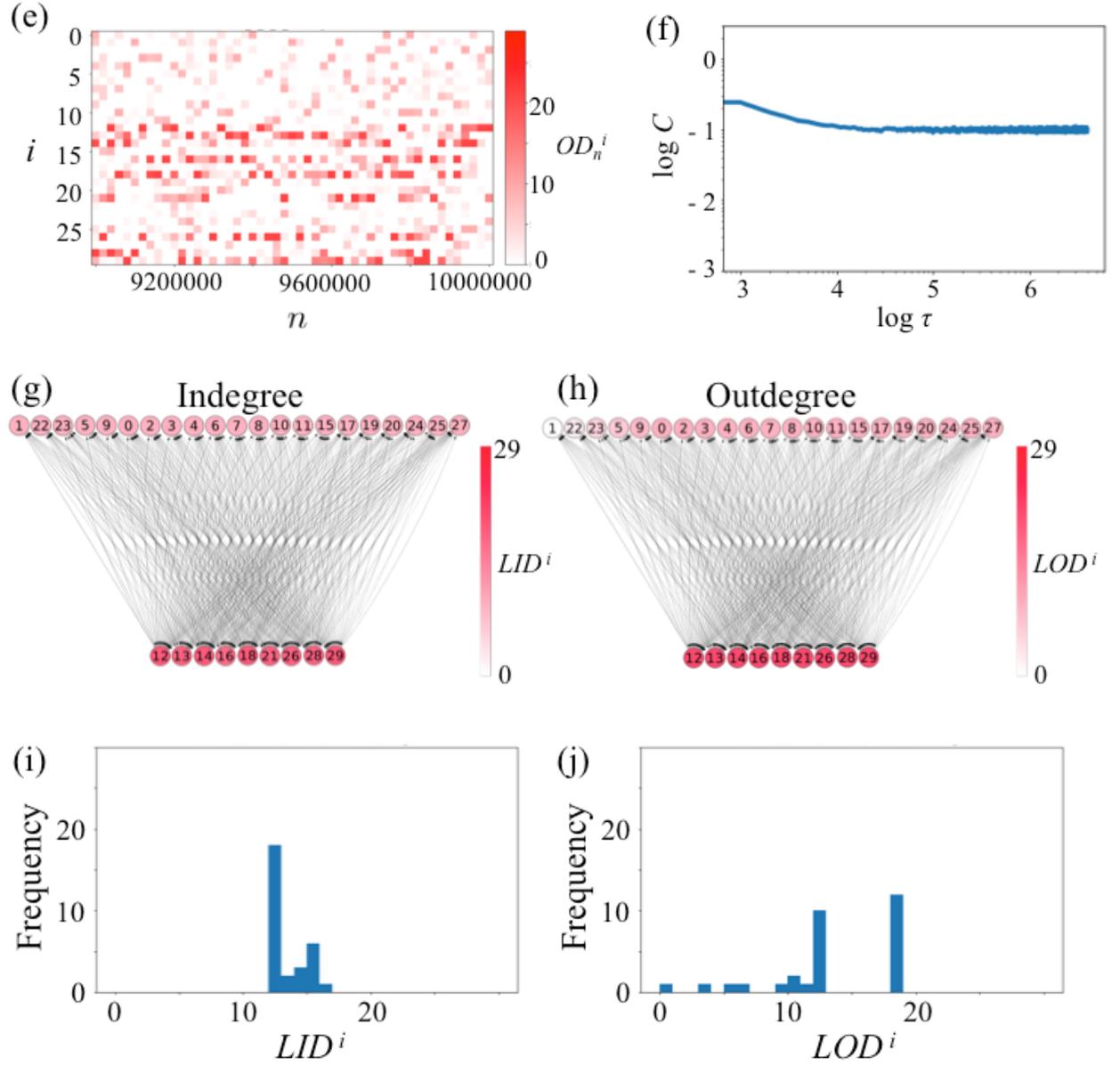

FIG. 9. (a-c) Example of a typical network structure, (a) $ID_n^i$, (b) $OD_n^i$ and (c) $x_n^i$ at $n = 9\times10^6$ and $n = 10^7$ of each element in the present model with $a = 3.75, c = 0.3, \delta = 1$. Here, the color of each circle indicates the values of (a) $ID_n^i$, (b) $OD_n^i$ and (c) $x_n^i$, and $i$-th element is connected from the $j$-th element by an arrow when $w_n^{ij} > \frac{1}{N-1}$. (d) Typical temporal evolutions of $x_n^i$ for $n = 9999950 \sim 10000000$, (e) typical temporal evolutions of $OD_n^i$ for $n = 9000000 \sim 10000000$, and (f) autocorrelation functions of connection strengths, $C(\tau)$, of the presented state. (g-h) Long time-averaged network structure, (g) $LID^i$, and (f) $LOD^i$ of the presented state. Here, the color of each circle indicates the values of (g) $LID^i$ and (f) $LOD^i$. When $\langle w_n^{ij} \rangle > \frac{1}{N-1}$ at $n = 10^7$, the $i$-th

element is connected from $j$-th element by an arrow. In this case, a dynamic network with "hidden paired layers network" was observed.

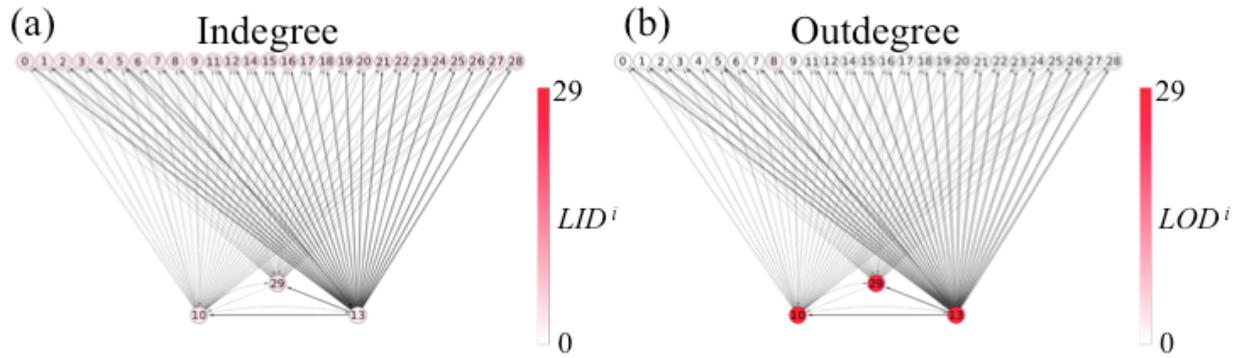

FIG. 10. Typical long time-averaged network structure, (a) $LID^i$, and (b) $LOD^i$ in the present model with $a = 4.0, c = 0.5, \delta = 1.0$. The meanings of the circles and arrows are the same as those in FIG. 9 (g-h). In this case, a dynamic network with "hidden paired layers network" involving the similar shape to the pacemaker network was observed.

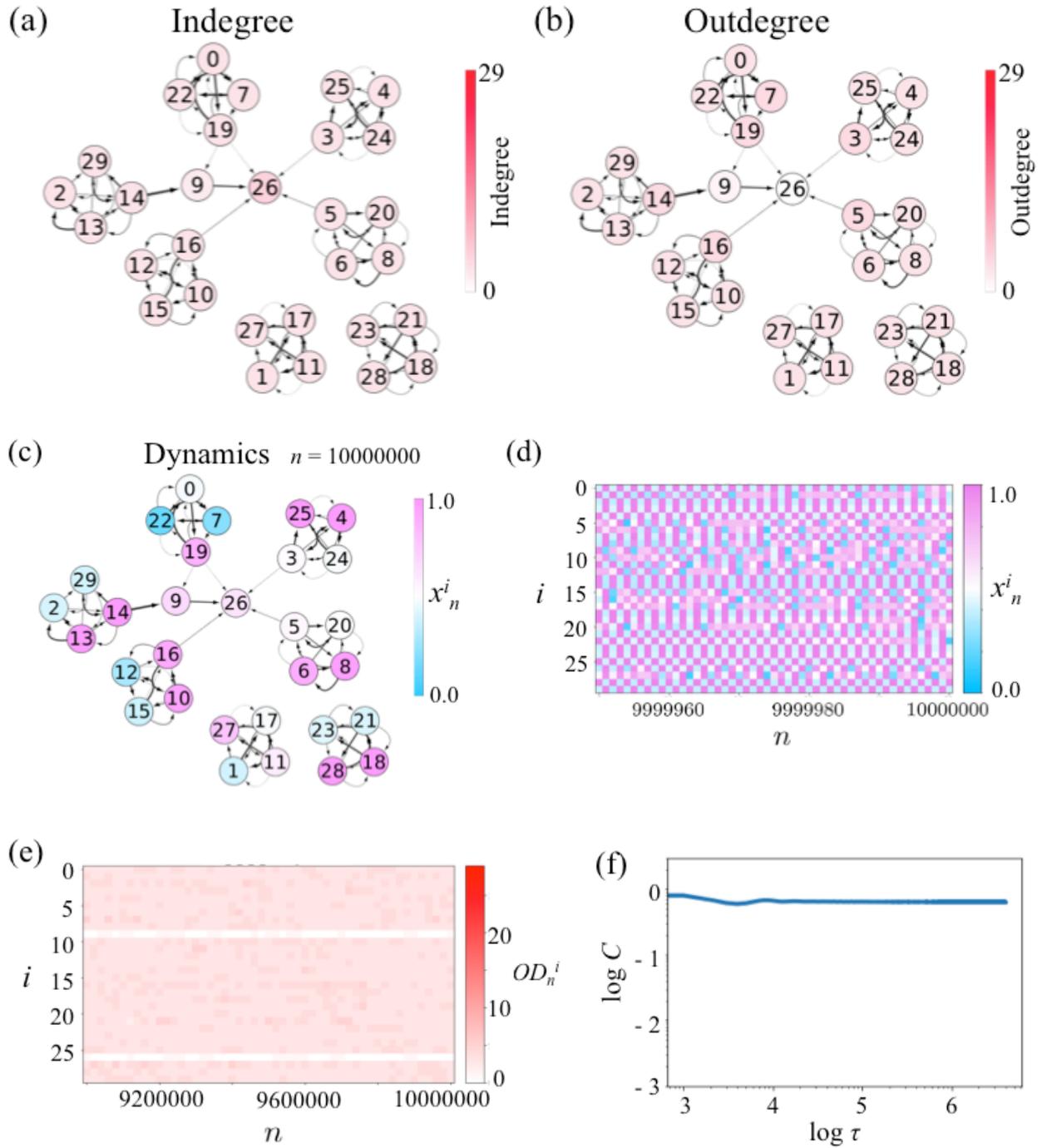

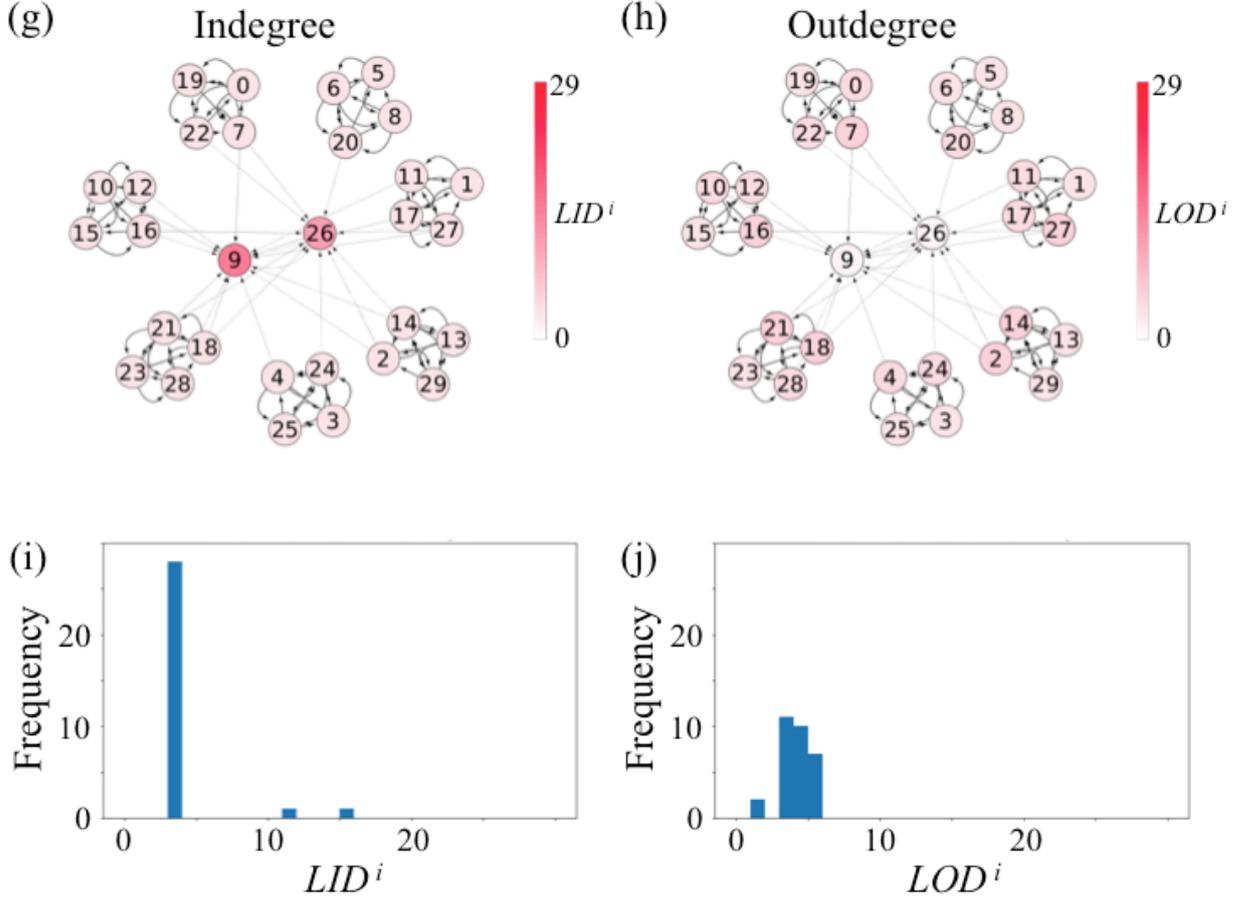

FIG. 11. (a-c) Example of a typical network structure, (a) $ID_n^i$, (b) $OD_n^i$, and (c) $x_n^i$ at $n = 10^7$ of each element in the present model with $a = 3.9, c = 0.2, \delta = 0.01$ Here, the meanings of circles and arrows are the same as those in FIG. 9 (a-c). (d) Typical temporal evolutions of $x_n^i$ for $n = 9999950 \sim 10000000$, (e) typical temporal evolutions of $OD_n^i$ for $n = 9000000 \sim 10000000$, and (f) autocorrelation functions of connection strengths $C(\tau)$, of the presented state. (g-h) Long time-averaged network structure, (g) $LID^i$, and (f) $LOD^i$ of the presented state. Here, the meanings of the circles and arrows are the same as those in FIG. 9 (g-h). In this case, a dynamic network with "hidden modular network" was observed.

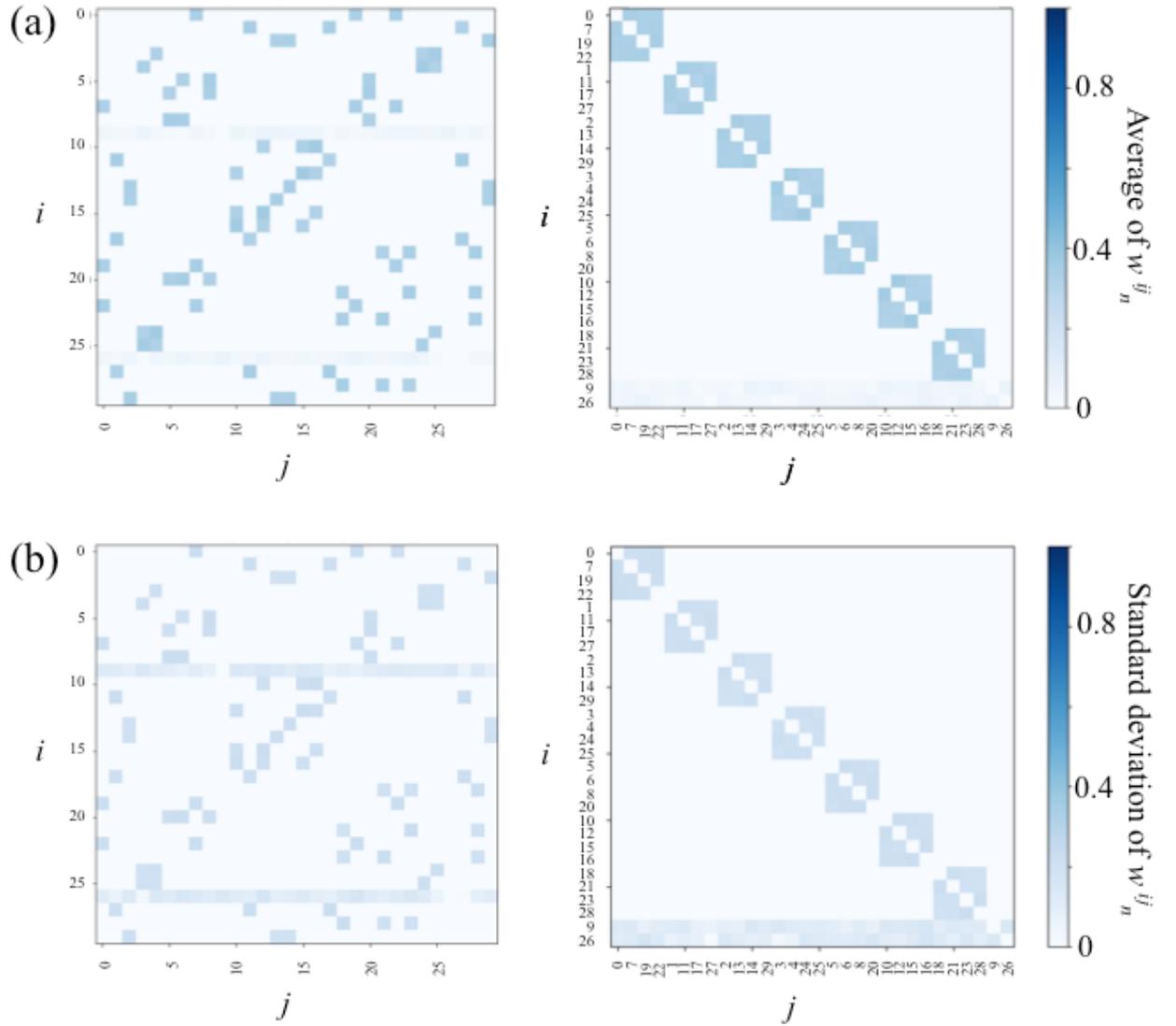

FIG. 12. (a) Average and (b) standard deviation of $w_n^{ij}$ over $n = 8\times10^6 \sim 10^7$ in the case that was introduced in FIG. 10, where the left indicates $w_n^{ij}$, in the order from small to large $i,j$, and right indicates $w_n^{ij}$ for the reordered $i,j$, which shows the connections among the four elements in each module clearly.

■ $a = 4, c = 0.2, \delta = 1$

(a) Indegree

(b) Outdegree

$n = 9000000$

$\Downarrow$ $n = 10000000$

$ID_n^i$

$OD_n^i$

(c) Dynamics

$n = 9000000$

$\Downarrow$ $n = 10000000$

$x_n^i$

(d)

$x_n^i$

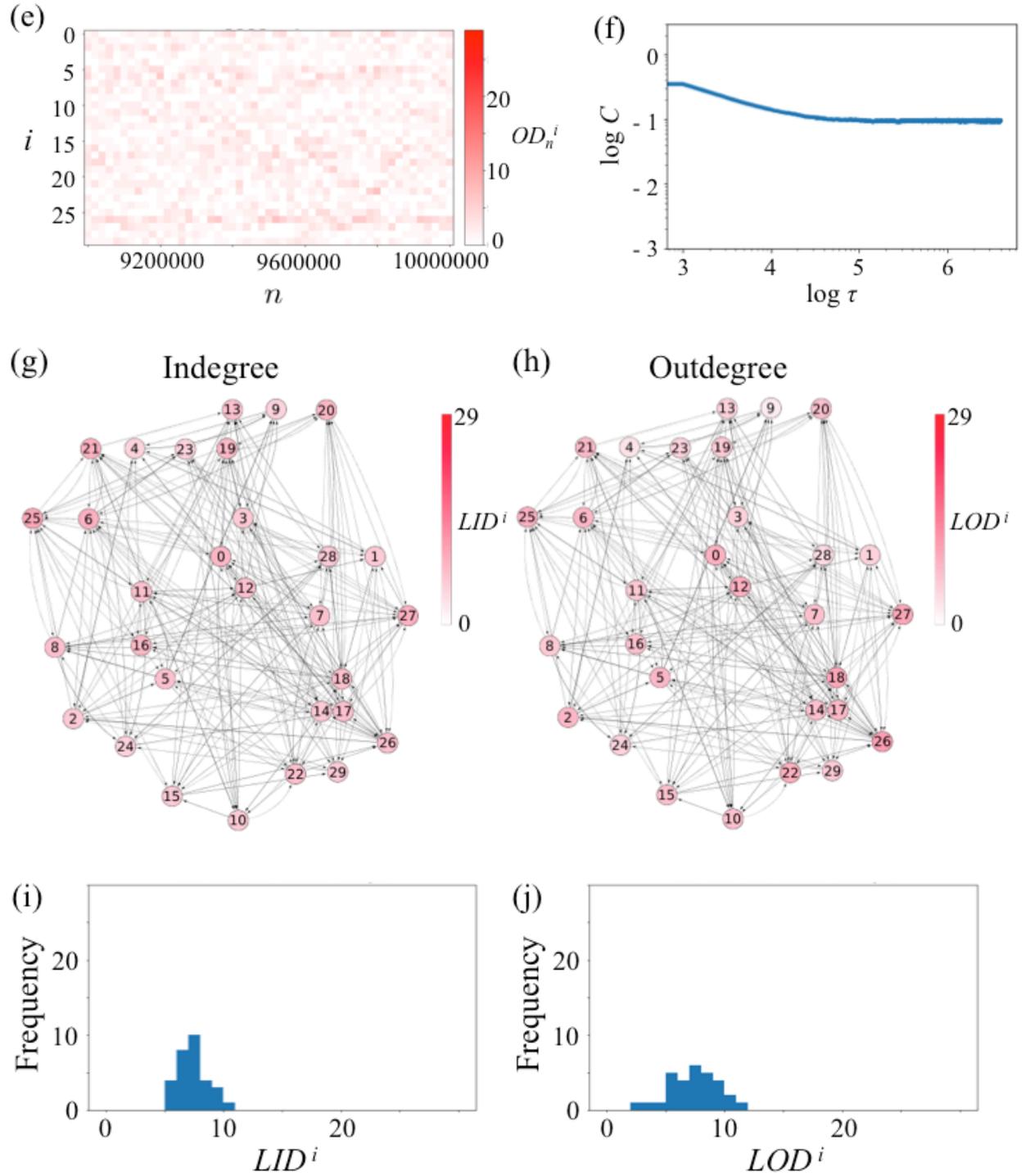

FIG. 13. (a-c) Example of typical network structure, (a) $ID_n^i$, (b) $OD_n^i$, and (c) $x_n^i$ at $n = 10^7$ of each element in the present model with $a = 4.0, c = 0.2, \delta = 0.01$ Here, the meanings of the circles and arrows are the same as those in FIG. 9 (a-c). (d) Typical temporal evolutions of $x_n^i$ for $n = 9999950 \sim 10000000$, (e) typical temporal evolutions of $OD_n^i$ for $n = 9000000 \sim 10000000$, and (f) autocorrelation functions of connection

strengths $C(\tau)$, of the presented state. (g-h) Long time-averaged network structure, (g) $LID^i$, (f) $LOD^i$, and histograms of (i) $LID^i$ and (j) $LOD^i$ of the presented state. Here, the meanings of the circles and arrows are the same as those in FIG. 9 (g-h). In this case, a dynamical network with a "hidden randomly connected network" was observed.

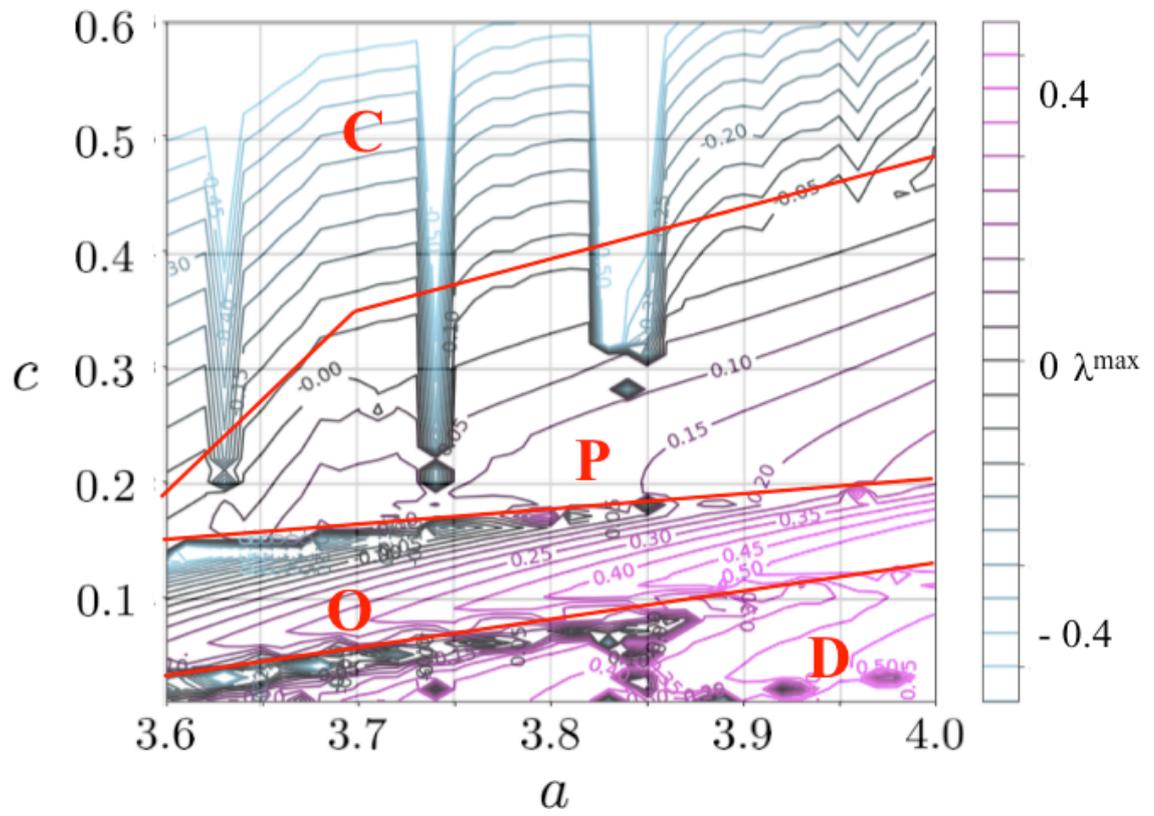

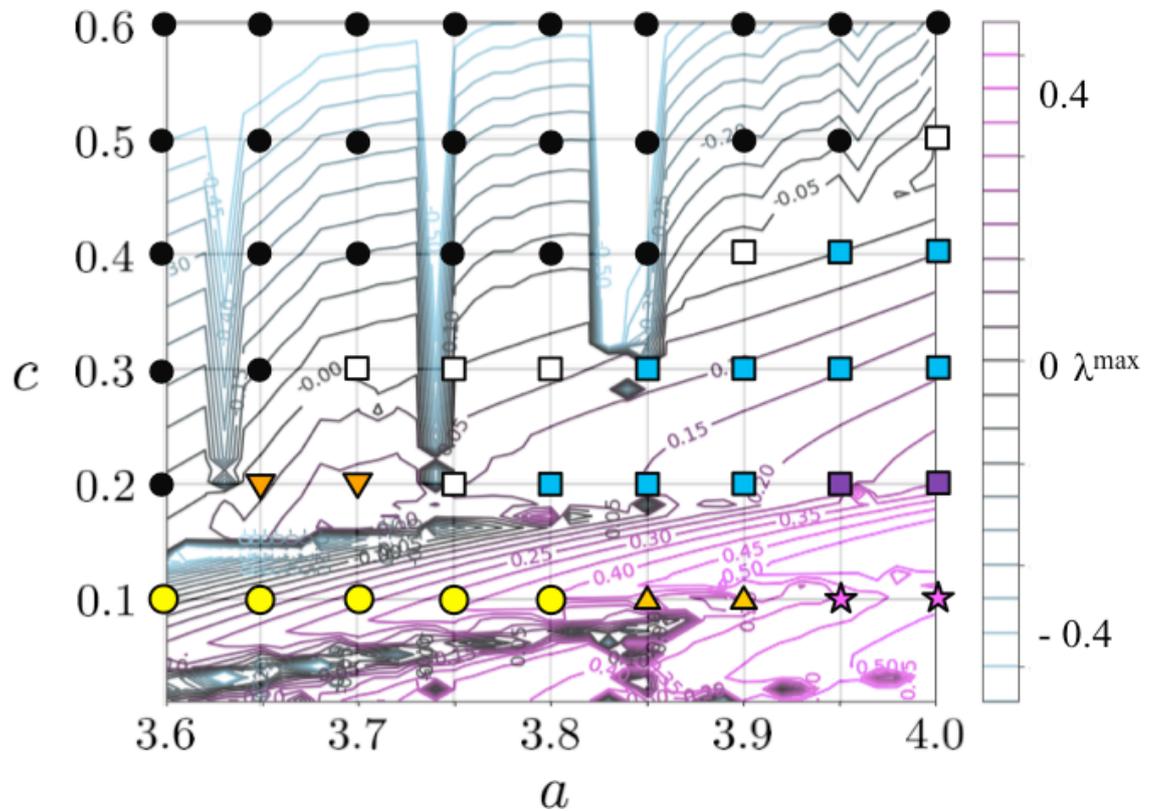

FIG. 14. (a) Maximum split exponents $\lambda^{max}$ of $\{x_n^i\}_i$ for $a$ and $c$ at $\delta = 1.0$. Red curves indicate the boundary of phases of phase diagram of dynamics of $\{x_n^i\}_i$ named coherent phase (C), ordered phase (O), partially ordered phase (P), and desynchronized phase (D) reported by recent research [21]. (b) The superposition of $\lambda^{max}$ and the phase diagram of the network structure was obtained in FIG. 1(a).

(a) $\delta = 0.1$

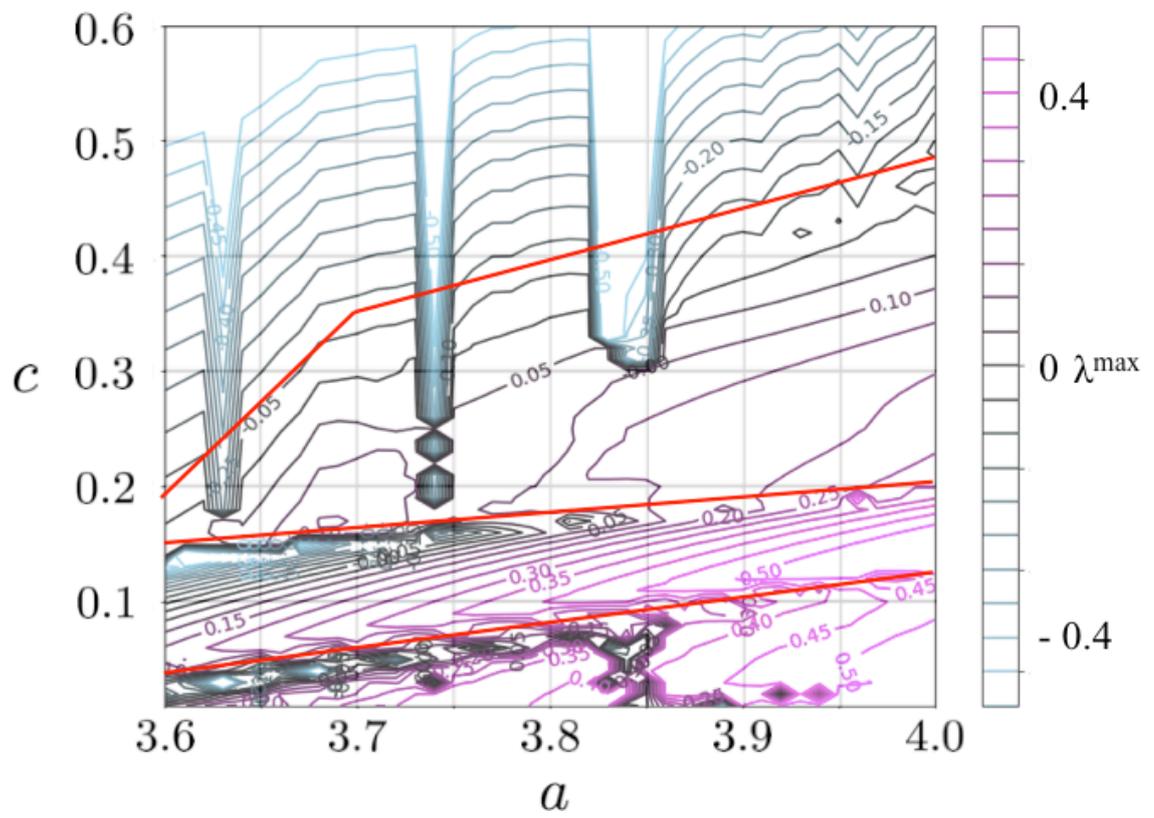

(b) $\delta = 0.01$

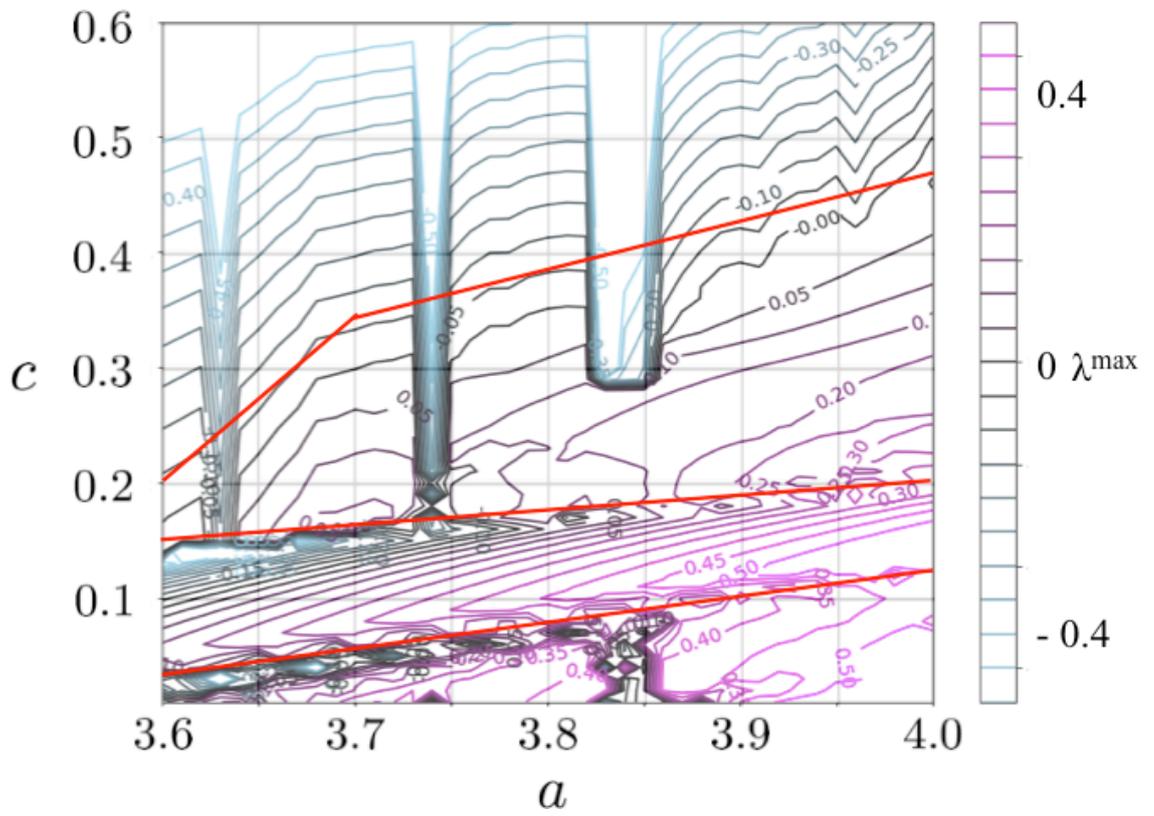

(c) $\delta = 0.1$

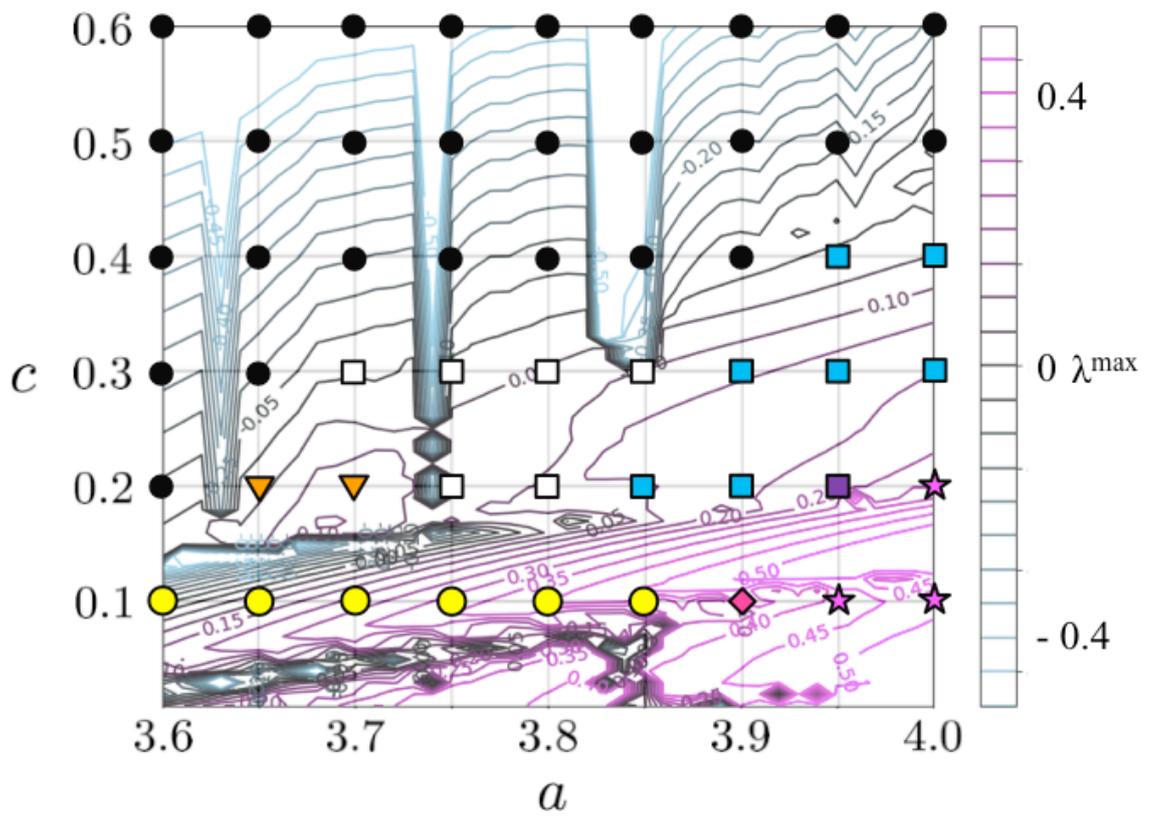

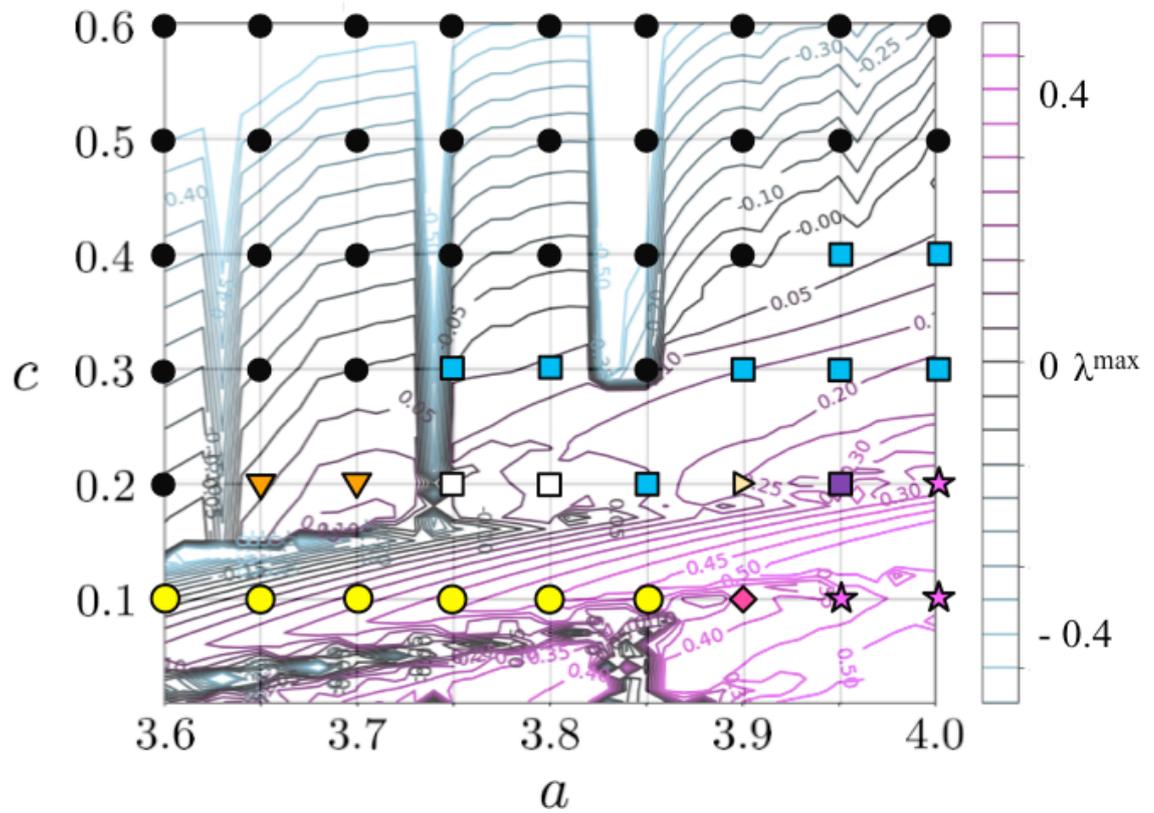

FIG. 15. (a-b) Maximum split exponents $\lambda^{max}$ of $\{x_n^i\}_i$ for $a$ and $c$ at (a) $\delta = 0.1$ and (b) $\delta = 0.01$. Each red curve indicates the expected one corresponding to that in FIG. 14 (a), respectively. (c-d) Superposition of $\lambda^{max}$ and the phase diagram of the network structure for (c) $\delta = 0.1$ and (d) $\delta = 0.01$ as obtained in FIG. 1(b) and (c), respectively.

# Appendix

(a) $\delta = 1$

Figure: Grid of log-log plots of $\log C$ vs $\log \tau$ for varying parameters $a$ (columns: 3.6, 3.65, 3.7, 3.75, 3.8, 3.85, 3.9, 3.95, 4.0) and $c$ (rows: 0.1, 0.2, 0.3, 0.4, 0.5).

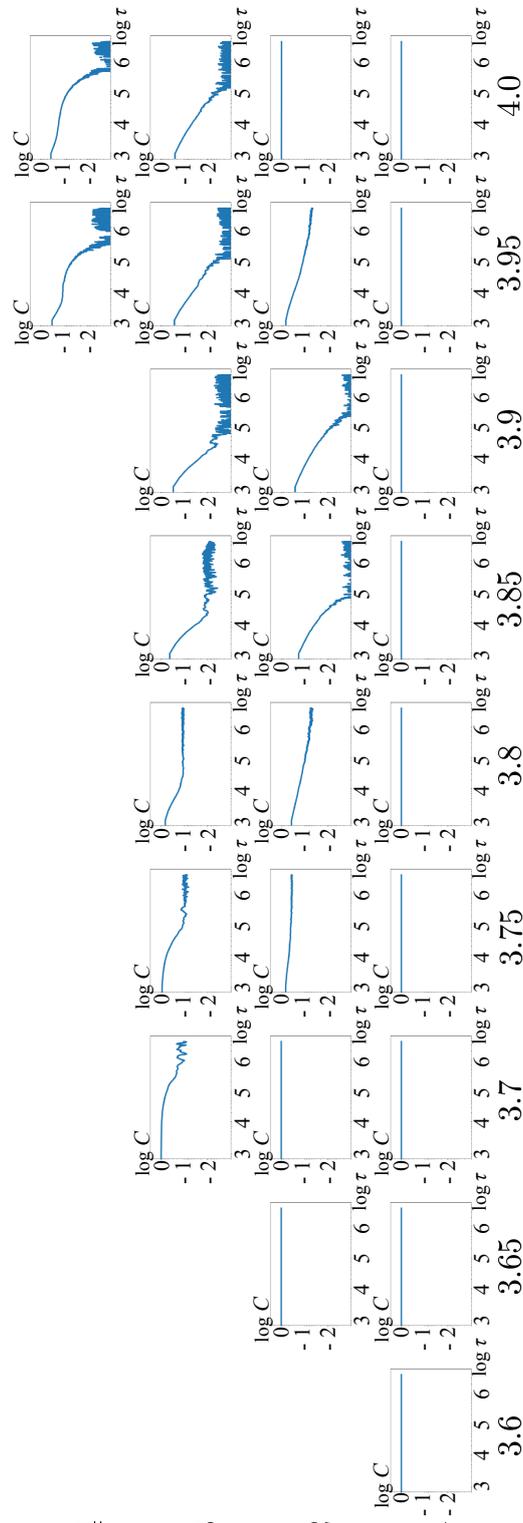

(b) $\delta = 0.1$

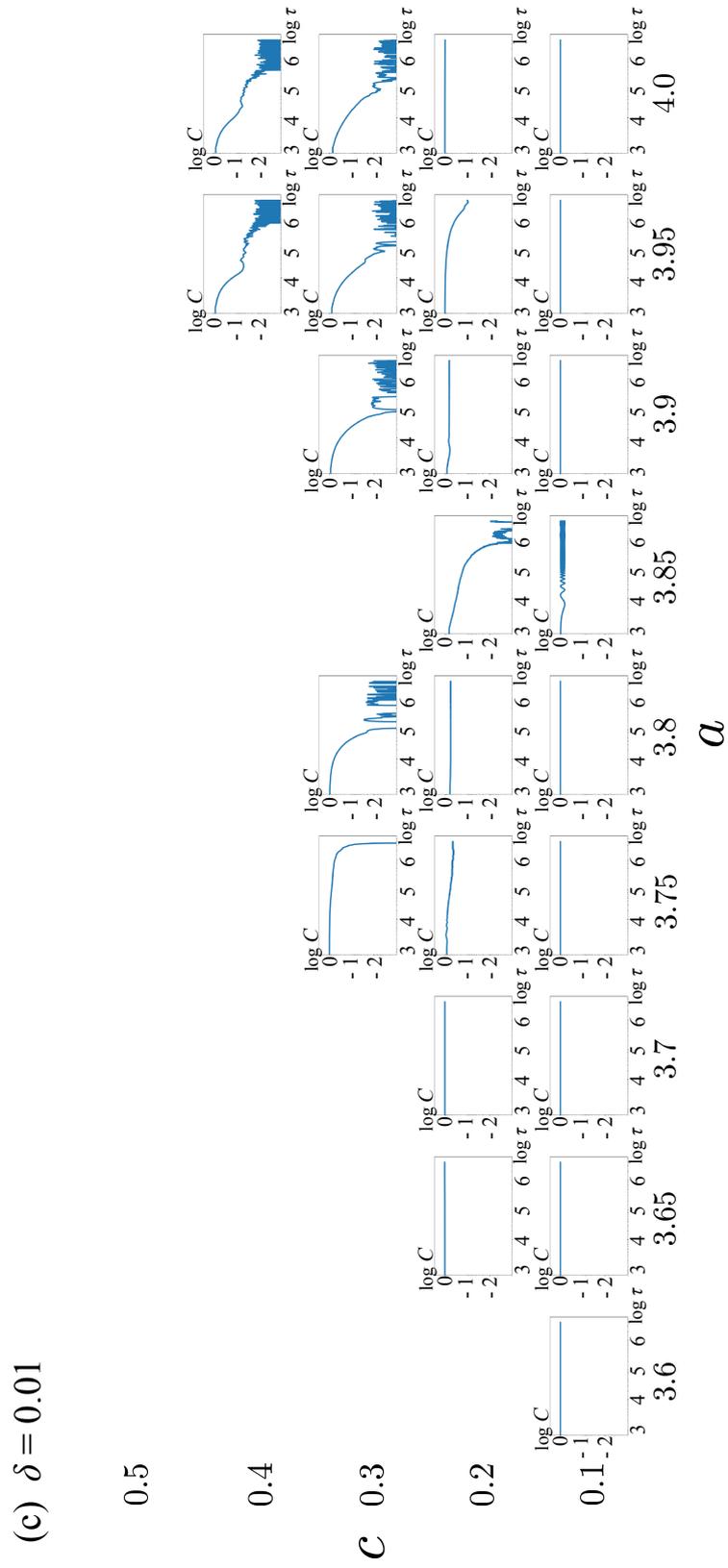

FIG. A1. Autocorrelation functions of connection strengths $C(\tau)$, for various $a$, and $c$ with (a) $\delta = 1.0$, (b) $\delta = 0.1$, and (c) $\delta = 0.01$.

[1] M. Anghel, Z. Toroczkai, K. E. Bassler, and G. Korniss, Phys. Rev. Lett. **92**, 4 (2004).

[2] N. Guttenberg and N. Goldenfeld, Phys. Rev. E **81**, 046111 (2010).

[3] C. Furusawa and K. Kaneko, Bull. Math. Biol. **60**, 659 (1998).

[4] A. Awazu and K. Kaneko, Phys. Rev. E **76**, 041915 (2007).

[5] A. Awazu and K. Kaneko, Phys. Rev. E **80**, 010902 (2009).

[6] A. Awazu and K. Kaneko, Phys. Rev. E **80**, 041931 (2009).

[7] K. Tokita and A. Yasutomi, Theor. Popul. Biol. **63**, 131 (2003).

[8] T. Shimada, S. Yukawa, and N. Ito, Artif. Life Robot. **6**, 78 (2002).

[9] J. Ito and K. Kaneko, Neural Networks **13**, 275 (2000).

[10] J. Ito and K. Kaneko, Phys. Rev. Lett. **88**, 287011, (2002).

[11] M. Funabashi, Int. J. Bifurc. Chaos **25**, 1550054. (2015).

[12] W. J. Yuan and C. Zhou, Phys. Rev. E **84**, 016116 (2011).

[13] J. Jost and K. M. Kolwankar, Phys. A Stat. Mech. Its Appl. **388**, 1959 (2009).

[14] O. V. Maslennikov and V. I. Nekorkin, Uspekhi Fiz. Nauk **187**, 745 (2017).

[15] S. Bornholdt and T. Röhl, Phys. Rev. E **67**, 066118 (2003).

[16] M. Girardi-Schappo, M. H. R. Tragtenberg, and O. Kinouchi, J. Neurosci. Methods **220**, 116, (2013).

[17] T. Mo, W. Kejun, Z. Jianmin and Z. Liying, Handb. Res. Artif. Immune Syst. Nat. Comput. Appl. Complex Adapt. Technol. **144**, 520, (2009).

[18] L. Chen and K. Aihara, Neural Networks **8**, 915 (1995).

[19] M. Adachi and K. Aihara, Neural Networks **10**, 83 (1997).

[20] Y. Hayakawa and Y. Sawada, Phys. Rev. E **61**, 5091 (2000).

[21] J. Ito and T. Ohira, Phys. Rev. E **64**, 066205 (2001).

[22] J. Ito and K. Kaneko, Underst. Complex Syst. **2009**, 137 (2009).

[23] A. E. Allahverdyan and A. Galstyan, PLoS One **11**, 0159301 (2016).

[24] V. Botella-Soler and P. Glendinning, Phys. Rev. E **89**, 062809 (2014).

[25] V. M. Eguíluz, M. G. Zimmermann, C. J. Cela-Conde, and M. San Miguel, Am. J. Sociol. **110**, 977 (2005).

[26] A. Herz, B. Sulzer, R. Kühn, and J. L. van Hemmen, Biol. Cybern. **60**, 457 (1989).

[27] K. Kaneko. Physica D **41**, 137 (1990)

[28] Y. Dan and M. Poo, Neuron **44**, 23 (2004).

[29] V. Botella-Soler and P. Glendinning, Euro. Phys. Lett. **97**, 50004, (2012).